%% file: main.tex
\documentclass[aps,prx,reprint,superscriptaddress,longbibliography,floatfix]{revtex4-2}
\usepackage{newtxtext,newtxmath,bm}
\usepackage{booktabs,array,tabularx}
\usepackage{xcolor}
\usepackage{graphicx}
\usepackage{tikz}
\usepackage{pgfplots}
\usepackage{hyperref}
\usetikzlibrary{arrows.meta,positioning,fit,calc,backgrounds}
\pgfplotsset{compat=1.18}

\newcommand{\Sep}{\mathrm{Sep}}

\newcommand{\EcertTwo}{\ensuremath{\mathrm{E}_{2}}}
\newcommand{\EcertThree}{\ensuremath{\mathrm{E}_{3}}}
\newcommand{\EcertFour}{\ensuremath{\mathrm{E}_{4}}}
\newcommand{\Egt}{\ensuremath{\mathrm{E}_{>2}}}

\begin{document}

\title{AI-Driven Discovery of Entanglement Witness in Spin Hamiltonians}

\author{Jiwei Tian}
\altaffiliation{Equal contribution}
\affiliation{School of Physics, Peking University, Beijing 100084, China}

\author{Mouyang Cheng}
\altaffiliation{Equal contribution}
\email{vipandyc@mit.edu}
\affiliation{Center for Computational Science and Engineering, MIT, Cambridge, MA 02139, USA}
\affiliation{Department of Materials Science and Engineering, MIT, Cambridge, MA 02139, USA}

\author{Mingda Li}
\email{mingda@mit.edu}
\affiliation{Center for Computational Science and Engineering, MIT, Cambridge, MA 02139, USA}
\affiliation{Department of Nuclear Science and Engineering, MIT, Cambridge, MA 02139, USA}

\begin{abstract}
Entanglement is a crucial resource in quantum science and engineering, but detecting it in many-body states from experimentally accessible observables remains challenging. 
Here we introduce EntNet, an AI-driven framework to discover compact combinations of spin observables that accurately distinguish entangled from fully separable many-body mixed states in one-dimensional spin-$1/2$ chains governed by short-range Hamiltonians.
We construct local-unitary-invariant representations of measurable Pauli correlators up to four-body order, sparsely select 16 candidate invariant feature channels, and learn their spatial structure using a Transformer architecture with rotary positional embeddings. Using only expectation values of these low-order observables, EntNet achieves $>98\%$ recall on certified entangled states with near-zero false-positive rates on separable states. 
Remarkably, it exhibits transferability across system sizes and distinct Hamiltonian families, including frustrated and disordered spin models, with minimal loss of accuracy.
Beyond detection, EntNet provides physical insight into entanglement detection by revealing a selective spatial organization of pair correlations, while its finite-temperature decision boundary tracks the excitation gap of Jordan–Wigner quasiparticles.
These results establish a transferable and physically interpretable framework for data-driven discovery of entanglement witnesses from local correlation functions.
\end{abstract}

\maketitle

\section{Introduction}
Quantum entanglement is one of the defining features of quantum mechanics in many-body systems. Beyond its fundamental significance, entanglement serves as a key resource in quantum communication, quantum computation, and quantum cryptography protocols \cite{Horodecki2009,Bennett2000,Dur2007,Gisin2002,Plenio2007}. It is also deeply connected to the structure and complexity of quantum many-body states. In particular, entanglement entropy and the entanglement spectrum for pure states characterize entanglement between subsystems, revealing information that is not readily captured by conventional local observables. Their characteristic behavior can provide signatures of quantum criticality, symmetry breaking, and topological order, making entanglement a powerful tool for characterizing strongly correlated quantum matter \cite{Amico2008,friis2019entanglement,Laflorencie2016}.
The scaling and structure of entanglement also strongly constrain the efficiency of tensor-network representations, such as matrix-product states (MPS) and projected entangled-pair states (PEPS), and help explain the success of numerical methods such as the density-matrix renormalization group (DMRG) \cite{White1992,Schollwock2011,Cirac2021}.
Reliable characterization of entanglement is therefore important both for understanding quantum matter and for developing efficient methods to represent and manipulate many-body quantum states. 

However, detecting entanglement in generic many-body mixed states is computationally challenging. In fact, even in a simplified case, such as the weak-membership problem for bipartite separability, the computational task is still NP-hard \cite{Gurvits2003}.
A variety of computable entanglement criteria and witnesses have consequently been developed, including the positive-partial-transpose (PPT) criterion and negativity, realignment and cross-norm criteria (CCNR) \cite{Horodecki2009,GuhneToth2009}, spin-squeezing inequalities \cite{toth2007optimal}, covariance-matrix criteria \cite{guhne2007covariance}, and the quantum-Fisher-information-based criteria \cite{hyllus2012fisher,Hauke2016}.
Recent advances have connected quantum Fisher information to dynamical susceptibilities and demonstrated entanglement witnessing using inelastic neutron scattering \cite{Hauke2016,Scheie2021}. Related protocols have also been proposed for resonant inelastic x-ray scattering (RIXS) to probe electronic and spin-orbital entanglement \cite{Liu2025RIXS,Shen2026RIXS}.
A three-spin chirality operator has also been proposed to characterize topological phases of Kitaev quantum spin liquids \cite{hwang2022identification}.
Despite these advances, existing approaches often require density-matrix information or rely on predefined witness constructions and specific experimentally accessible observables, potentially overlooking broader entanglement signatures.
This motivates the search for generalizable entanglement signatures that can be learned from diverse low-order local spin correlations without reconstructing the full quantum state, which is generally inaccessible in condensed matter systems.

Recent advances in artificial intelligience (AI) and machine learning (ML) provide a complementary, data-driven route to identifying entanglement from experimentally or numerically accessible information \cite{Lu2018,Chen2022}. 
ML models have been used, for example, to infer logarithmic negativity or entanglement entropy from a restricted set of observables \cite{Gray2018,Berkovits2018}, or to classify entanglement within structured families such as few-body GHZ- and W-type states \cite{Chen2021,Urena2024}. 
Two key challenges nevertheless remain. First, extending such approaches to spatially extended many-body systems requires capturing correlations across different length scales while remaining transferable across system sizes and Hamiltonians. 
Second, practical entanglement detection calls for methods that operate directly on experimentally accessible observables without reconstructing the full density matrix or assuming a narrowly prescribed family of target states.
Addressing these limitations in extended many-body systems demands a scalable and transferable ML framework capable of resolving entanglement across diverse states, system sizes, and Hamiltonians.

In this work, we develop EntNet, an ML framework for detecting entanglement in finite-temperature mixed states of extended one-dimensional spin-$1/2$ systems using \emph{only} low-order spin correlation functions. A central consideration is that entanglement is invariant under local unitary (LU) transformations of individual spins, whereas the measured spin correlations generally depend on the choice of local basis. Learning this invariance directly from data is difficult, particularly when a large number of potentially redundant correlators are involved. To address these challenges, we construct LU-invariant representations of two-, three-, and four-body correlators using singular values and tensor contractions, and sparsely distill high-dimensional candidate invariant channels into only 16 features. Their spatial organization is subsequently learned by a Transformer architecture with rotary positional embeddings (RoPE) \cite{Vaswani2017,Su2024}.
Trained using 35,000 synthetically generated $N=16$ spin-chain states, comprising certified entangled and fully separable examples, EntNet achieves $>98\%$ detection accuracy on the evaluated benchmarks, while maintaining near-zero false-positive rates on separable states.
Without retraining, the model transfers from $N=16$ to $N=32$ $XY$ chains and generalizes across the tested $XXZ$, frustrated $J_1$-$J_2$, and random local spin Hamiltonians, with minimal loss of accuracy.
Beyond the high detection accuracy, the distilled representation from EntNet provides physical insight into the entanglement structure encoded in low-order observables. 
For example, pair correlations at lattice-site separations $r=1$ and $r=4$ emerge as particularly important, in contrast to the negligible contributions from $r=2$ and $r=3$ pair features.
At finite temperature, EntNet further resolves a narrow crossover between entangled and separable regimes: in the saturated XX chain, its lower decision boundary quantitatively tracks the thermal excitation scale set by the quasiparticle gap.
These results establish EntNet as a scalable AI-driven framework for discovering compact, transferable, and physically interpretable entanglement witnesses from low-order observables in generic many-body mixed states.

\section{Method}
EntNet detects many-body entanglement from low-order Pauli correlators without requiring full quantum-state tomography. We first construct labeled spin-chain states and extract LU-invariant features from local spin correlators.
Sparse feature selection then reduces the candidate observables to a compact set of 16 channels, which form site-resolved inputs to a Transformer that captures their spatial organization. Below, we describe the construction of training labels and datasets, the invariant representation, and the architecture of EntNet, which are also illustrated in Fig.\,\ref{fig1}.

\subsection{Entanglement and computable certificates}

For a system of $N$ spins, a pure state is fully separable if it is a product state $|\psi\rangle=\bigotimes_{i=1}^{N}|\phi_i\rangle$. A mixed state is fully separable if it can be written as a probabilistic mixture of product states,
\begin{equation}
    \rho_{\Sep}=\sum_\mu p_\mu\bigotimes_{i=1}^{N}\rho_i^{(\mu)},
    \qquad p_\mu\geq0,\quad\sum_\mu p_\mu=1.
\label{eq:intro-separable}
\end{equation}
States that do not admit such a decomposition are entangled \cite{Horodecki2009}. Rather than attempting the generally difficult separability test for the full many-body state, we search for entanglement in small subsystems. 
For each subset $S$ of $k=2,3,$ or $4$ sites, we construct its reduced density matrix $\rho_S=\mathrm{Tr}_{\bar S}\rho$, where $\rho$ describes the full $N$-spin system and $\mathrm{Tr}_{\bar S}$ traces out the remaining $N-k$ spins, leaving the quantum state of the selected subsystem $S$.
Because every marginal of a fully separable state must itself be separable, finding an entangled marginal is sufficient to certify entanglement of the full $N$-site state.

For each reduced density matrix $\rho_S$, we test every nontrivial bipartition $A|B$ of the selected sites using two complementary separability criteria: PPT and CCNR \cite{Horodecki2009,GuhneToth2009}.
For PPT, we partially transpose the density matrix with respect to subsystem $A$, denoted by $\rho^{T_A}$, whose matrix elements in a product basis satisfy $(\rho_S^{T_A})_{i\alpha,j\beta}=(\rho_S)_{j\alpha,i\beta}$. 
For CCNR, we realign the density matrix across the same bipartition, denoted by $\mathcal{R}_{A|B}(\rho)$, by rearranging its matrix elements as
$[\mathcal{R}_{A|B}(\rho_S)]_{ij,\alpha\beta}=(\rho_S)_{i\alpha,j\beta}$, and compute the trace norm, i.e. the sum of its singular values. Here, $i,j$ and $\alpha,\beta$ index basis states of subsystems $A$ and $B$, respectively.
For separable states, which are convex mixtures of product states, partial transposition preserves the positivity of each product-state contribution, while realignment yields a trace norm no greater than unity.
Therefore, any state separable across $A|B$ must satisfy
\begin{equation}
    \lambda_{\min}\!\left(\rho_S^{T_A}\right)\geq 0,
    \qquad
    \left\|\mathcal{R}_{A|B}(\rho_S)\right\|_1\leq 1.
\end{equation}
Here, $\lambda_{\min}$ is the smallest eigenvalue and $\|\cdot\|_1$ is the trace norm. 
Thus, either a negative eigenvalue of \(\rho_S^{T_A}\) or a realignment norm exceeding unity rigorously certifies entanglement across that bipartition \cite{Peres1996,Horodecki1996,Rudolph2005,ChenWu2003}.

We use these sufficient criteria to assign entanglement labels \cite{Lu2018,Urena2024}, testing subsystems in increasing size. A state is labeled \EcertTwo{} if any two-site marginal violates PPT; otherwise, a PPT or CCNR violation in a three-site marginal assigns \EcertThree{}, and, if no lower-order certificate is found, a violation in a four-site marginal assigns \EcertFour{}. These labels indicate the smallest subsystem size at which our chosen criteria certify entanglement.
All \EcertTwo{}, \EcertThree{}, and \EcertFour{} states could enter the dataset as positive (entangled) examples.
Negative examples are separable states generated by construction from product states and their convex mixtures, as defined in Eq.~\eqref{eq:intro-separable} (Fig.\,\ref{fig1}b). States with no detected certificate are labeled ``unknown'' and excluded from EntNet training and testing to avoid ambiguous supervision \cite{PawlowskiKrawczyk2024}. Further details are given in Appendix~A.

\subsection{Dataset preparation}
The core training dataset consists of 35,000 synthetically generated states of $N=16$ spin-$1/2$ systems, equally divided between separable and certified entangled states.
To evaluate the model on physically motivated quantum states, we primarily consider the finite-temperature Gibbs states of the anisotropic XY spin chain with periodic boundary condition during evaluation, as shown in Fig.\,\ref{fig1}a,
\begin{equation}
H_{\rm XY}=-\sum_{i=1}^{N}J\left[
\frac{1+\gamma}{2}X_iX_{i+1}+\frac{1-\gamma}{2}Y_iY_{i+1}\right]
-h\sum_{i=1}^{N}Z_i,
\label{eq:hamxy}
\end{equation}
where $i+N\equiv i$. The corresponding two-, three-, and four-body spin correlators can be evaluated using the Jordan--Wigner solution \cite{LiebSchultzMattis1961,BarouchMcCoyDresden1970,BarouchMcCoy1971}.
The negative class of the dataset is generated inside the fully separable set, including product states and convex mixtures of independently oriented product states according to Eq.\,\eqref{eq:intro-separable}. 
Candidate entangled states are first generated from random MPS \cite{Schollwock2011,Cirac2021}, then verified from \EcertTwo{}, \EcertThree{} and \EcertFour{} criteria.
We also test the transferability of EntNet to other 1D spin Hamiltonians, including open XXZ model, frustrated $J_1$-$J_2$ model and random local spin Hamiltonians. 
Further details on the target Hamiltonians and numerical protocols are shown in Appendix~A.

Additionally, we note that unless otherwise stated, the default training set includes only \EcertTwo{} states as entangled examples, while \EcertThree{} and \EcertFour{} states are deliberately held out for testing. 
This setup allows us to assess whether EntNet can generalize beyond the entanglement structures explicitly represented during training and recognize more complex entangled states from their low-order correlation signatures.

\subsection{LU-invariant representation}
The measured expectation values for Pauli spin correlators constitute the raw physical input to EntNet. Because entanglement is invariant under local basis rotations \cite{Horodecki2009,Plenio2007}, the density matrices $\rho$ and $U\rho U^\dagger$, with $U=\bigotimes_i U_i$, should receive the same ML label even though their raw Pauli correlators generally differ. Rather than learning this symmetry through data augmentation, EntNet imposes LU invariance directly through its feature input.

For sites $i_1<\cdots<i_q$, we define the $q$-body Pauli correlation tensor
\begin{equation}
T^{(q)}_{\alpha_1\cdots\alpha_q}=\mathrm{Tr}\!\left[\rho\,\sigma_{i_1}^{\alpha_1}\cdots\sigma_{i_q}^{\alpha_q}\right],
\end{equation}
where $\alpha_j\in\{x,y,z\}$ specifies the Pauli component at site $i_j$, with the one- through four-body sectors denoted $T^{(1)}$-$T^{(4)}$. Under a local unitary transformation, each Pauli index transforms through an orthogonal rotation $R_i\in\mathrm{SO}(3)$. Accordingly, after reshaping a correlation tensor across a bipartition $A|B$,
\begin{equation}
T^{(q)}_{A|B}\mapsto R_A T^{(q)}_{A|B}R_B^{\mathsf T},
\end{equation}
where $R_A$ and $R_B$ are Kronecker products of the corresponding local rotations. Because left- and right- multiplication by orthogonal matrices leaves singular values unchanged, the singular values of these tensor unfoldings provide exact LU-invariant features. 
EntNet uses these singular values, together with additional LU-invariant tensor norms and contractions, as its input representation, with further construction details given in Appendix~A, B.

\subsection{Transformer architecture and feature distillation}
\begin{figure*}[!t]
\centering
\includegraphics[width=\textwidth]{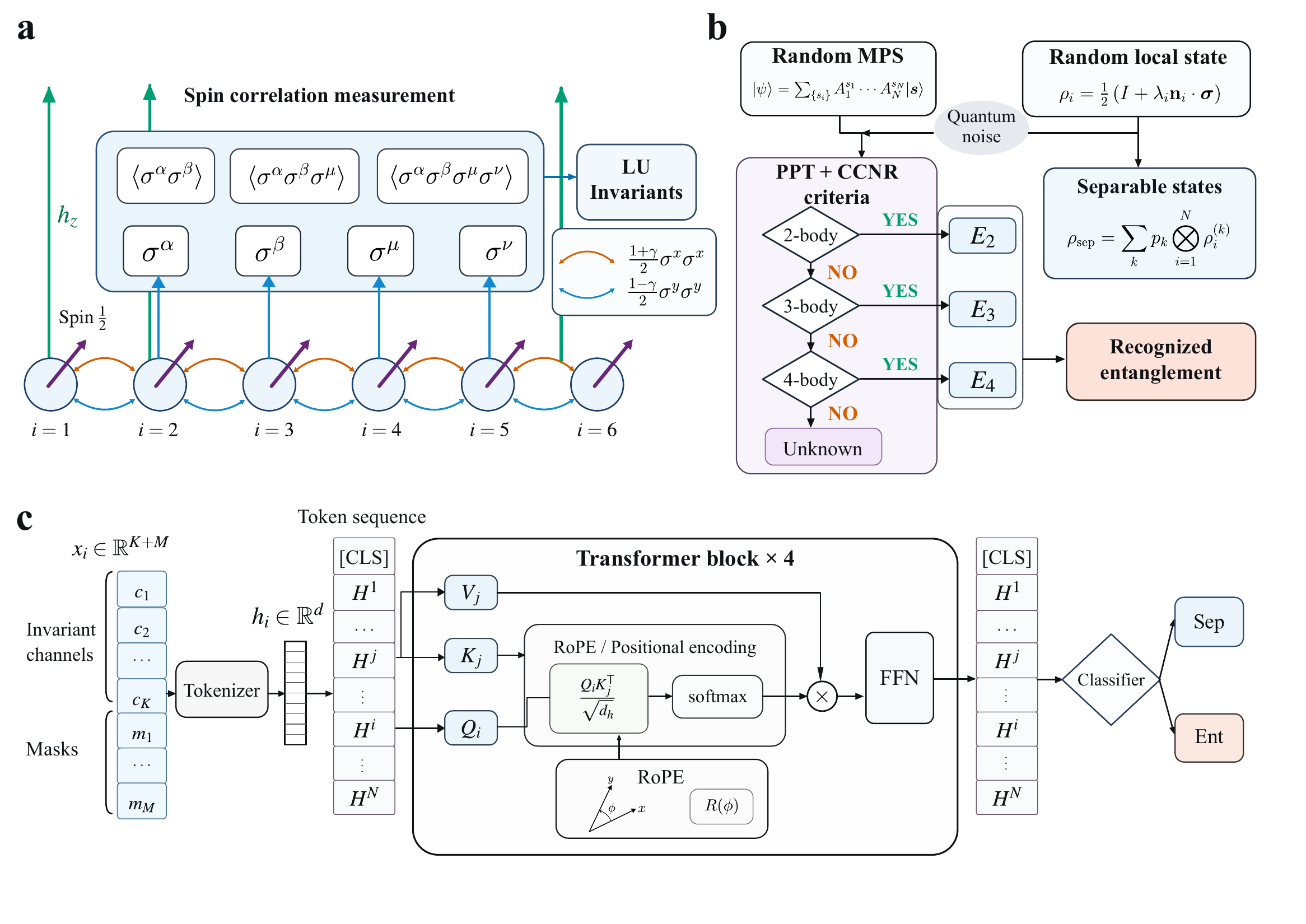}
\caption{Overview of the correlator-based entanglement detection framework. \textbf{a.} Schematic illustration of an XY spin chain under transverse field, where the low-order Pauli correlators are measured and their expectation values are used to construct local-unitary (LU) invariant features. \textbf{b.} Product states and their mixtures provide constructively separable samples, while locally noised random matrix-product states provide candidate entangled states. A hierarchy of two- to four-site PPT/CCNR tests assigns entanglement certificates (E2–E4) or leaves states uncertified. 
\textbf{c.} The selected K LU-invariant channels and geometry masks form site tokens, which are tokenized and then processed by a four-block Transformer with rotary positional encoding (RoPE). The final CLS representation is used for classification.}
\label{fig1}
\end{figure*}

As illustrated in Fig.\,\ref{fig1}c, EntNet organizes the LU-invariant correlation features into a sequence along the spin chain. At each lattice site $i$, the invariant channels are concatenated with geometry masks accounting for boundary effects. 
The resulting vector $\bm{x}^i$ is projected into a shared embedding $\bm{h}^i$, giving the token sequence
$$
[\mathrm{CLS}],\,\bm h^1,\ldots,\bm h^N .
$$
A four-layer Transformer \cite{Vaswani2017} then exchanges information between correlation patterns at different lattice positions through self-attention. The final CLS representation at the first token summarizes the sequence and is mapped to predict whether the probed quantum state is entangled or separable.

To encode how these local correlation patterns are arranged in space, we incorporate rotary position embedding (RoPE) \cite{Su2024} into the attention mechanism. As is shown in Fig.\,\ref{fig1}c, for each two-dimensional query-key component,
\begin{align}
\bm Q'_{i,\ell}
&=R(i\omega_\ell)\bm Q_{i,\ell},
&\bm K'_{j,\ell}
&=R(j\omega_\ell)\bm K_{j,\ell},
\end{align}
such that
\begin{equation}
(\bm Q'_{i,\ell})^{\mathsf T}\bm K'_{j,\ell}=\bm Q_{i,\ell}^{\mathsf T}R[(j-i)\omega_\ell]\bm K_{j,\ell}.
\label{eq:rope-relative-main}
\end{equation}
where \(R(\phi)\) is the two-dimensional rotation matrix
$$
R(\phi)=\begin{pmatrix}\cos\phi & -\sin\phi\\\sin\phi & \cos\phi\end{pmatrix},
$$
and $\omega_\ell$ is the rotation frequency associated with component $\ell$. 
The attention between two tokens therefore explicitly depends on their relative separation $j-i$. This is particularly useful for length extrapolation, where a relation learned between correlation patterns can be reused at the same separation in longer spin chains.

To identify which physical observables are most informative for entanglement detection, we place a learnable gate on each candidate LU-invariant channel, following the stochastic gating approach of Louizos \emph{et al.} \cite{Louizos2018}. 
Each gate acts on an entire observable channel and is shared across lattice positions. The gates are optimized jointly with EntNet while being encouraged to suppress channels that contribute little to classification. Their learned activation strengths therefore provide a direct measure of the relative importance of the different Pauli-correlation features used by the model. 
We then retain the top $K=16$ channels for subsequent analysis, yielding a lightweight distilled EntNet with $\sim$480,000 model parameters from an initial set of 117 candidate correlator channels, corresponding to an 86\% reduction in feature dimensionality.
Unlike generic learned latent features, these channels are explicitly defined LU-invariant functions of spin correlators, allowing the selected inputs to retain a physical interpretation. 
However, this reduction in feature dimensionality does not necessarily imply an equivalent reduction in the number of Pauli measurements, since constructing one invariant may require multiple correlators.
Finally, after training, EntNet classifies states as entangled or separable from the sigmoid output using a fixed decision threshold of $p=0.5$. Further architectural and evaluation details are provided in Appendix~C.

\section{Results}
\subsection{Learned compact representation of spatial correlation structure}
We first determine how many physical channels are required for reliable entanglement detection. Here, $K$ denotes the number of LU-invariant feature channels shared across lattice sites. As shown in Fig.\,\ref{fig:results}b, $K=16$ gives the best overall performance on the XY model, with all metrics approaching unity. 
With fewer channels, the distilled representation does not retain sufficient information for robust classification, most notably for the held-out $\mathrm{E}_{>2}$ states used to test extrapolation beyond the \EcertTwo{} entanglement class. 
Increasing $K$ beyond 16 provides no systematic improvement and eventually degrades performance, including a sharp loss of true-negative rate for the largest feature sets. 
These results indicate substantial redundancy among the candidate physical observables and motivate the use of the distilled 16-channel representation throughout the remainder of this work.

Having fixed the representation channels at $K=16$, we next examine which correlation features are retained by the distilled model.
The 16-channel representation contains five two-site features, including three of them at separation $r=1$ and two at $r=4$, together with contiguous three-site ($T^{(3)}$) and four-site ($T^{(4)}$) features (Fig.\,\ref{fig:results}a). 
Surprisingly, no explicit $r=2$ or $r=3$ pair channel is retained. Here, $r=|i-j|$ denotes the separation between two lattice sites; the absence of a selected channel at a given $r$ does not imply that the corresponding spin correlations vanish.
As shown in Fig.\,\ref{fig:results}b, the first $r=3$ and $r=2$ pair features enter only beyond the $K=16$ cutoff. 
Notably, this contrast in the contributions from different separations $r$ is obtained from training on a 16-site spin Hamiltonian, and whether it persists in the thermodynamic limit remains to be established.
The appearance of the $r=2$ channel further coincides with a pronounced degradation in the true-negative rate, suggesting that these intermediate-range pair features provide little additional discriminative information.
These results indicate that entanglement detection does not simply favor correlations according to spatial proximity. 
The $r=1$ pair features naturally capture the dominant nearest-neighbor correlations generated by local interactions, whereas the survival of $r=4$ features suggests that more extended two-site correlations provide complementary information not already contained in the local pair sector. 
The absence of explicit $r=2$ and $r=3$ pair channels reflects the learned compact representation, rather than the physical irrelevance of intermediate-range correlations.
Indeed, adding training examples of entangled states dominated by second-neighbor ($r=2$) correlations improves the $J_1$-$J_2$ \EcertTwo{} recall from 0.865 to 1.000 without changing the selected 16 channels, suggesting that the retained features encode relevant spatial information that appropriate training examples help the classifier exploit (Section III.B, Appendix~D).
Feature-ablation tests further confirm the complementary roles of these channels: higher-order features are essential for detecting higher-order entanglement, while additional four-body features cannot fully replace the selected longer-range pair correlations (Appendix~D).

\subsection{Generalizability of EntNet}
\begin{figure*}[t]
\centering
\includegraphics[width=\textwidth]{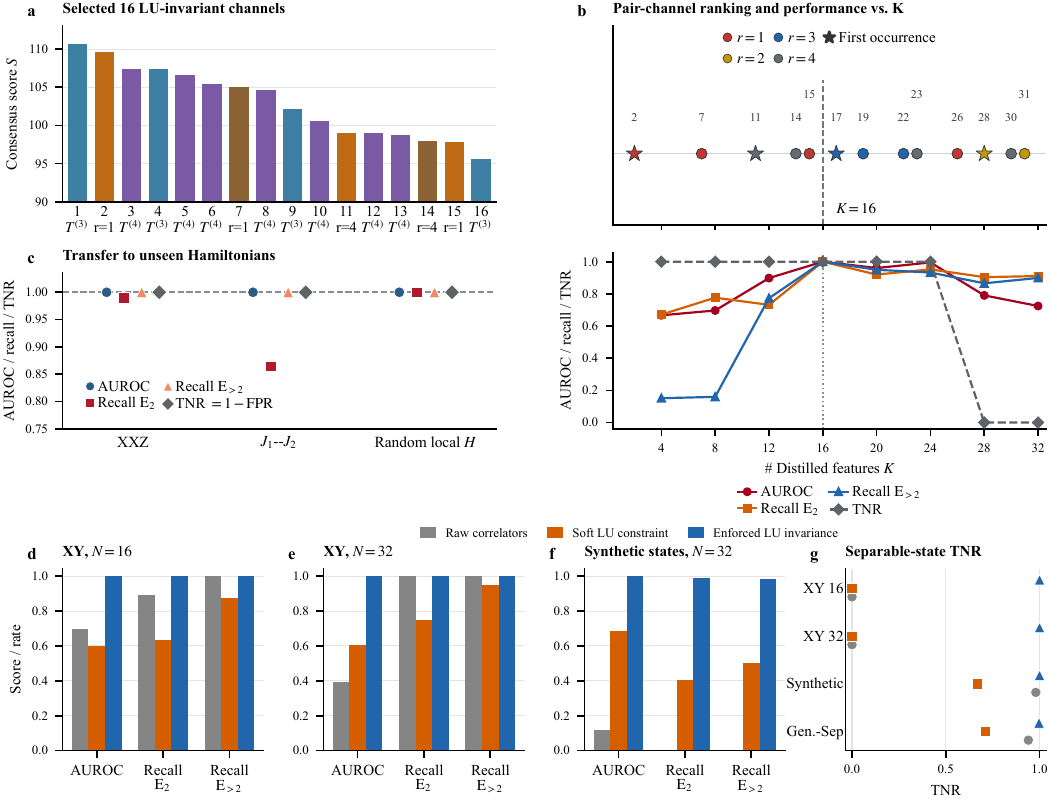}
\caption{
\textbf{Sparse LU-invariant feature selection and generalization of EntNet.}
\textbf{a.} The 16 selected local-unitary (LU) invariant channels, ordered by their consensus ranking across five independent feature selectors. The consensus score $S=118-\bar r$,
where $\bar r$ is the mean rank among 117 candidate channels, quantifies the preference for each feature.
\textbf{b.} Distribution of two-site channels in the consensus ranking (upper) and classification performance as a function of the number of selected channels $K$ (lower).
Stars mark the first occurrence of each pair separation, and the vertical dashed line indicates $K=16$. Classifiers in the lower panel are independently trained and evaluated on $N=16$ XY-chain states.
\textbf{c.} Generalization to XXZ, frustrated $J_1$--$J_2$, and random-local-Hamiltonian systems, none of which are included in training. The horizontal dashed line indicates unity.
\textbf{d--f.} Comparison of models using raw correlators, softly constrained LU representations, and exactly LU-invariant features on
\textbf{d.} $N=16$ XY-chain states, \textbf{e.} $N=32$ XY-chain states, and \textbf{f.} independently constructed $N=32$ states with diverse entanglement structures.
\textbf{g.} True-negative rates on the separable subsets of these benchmarks and an additional independent set of 2,000 constructively separable states.
All classifiers are trained only on synthetically generated $N=16$ separable and \EcertTwo{} states.
Here $E_{>2}$ includes \EcertThree{} and \EcertFour{} states. Recall and true-negative rate (TNR) are evaluated at the fixed threshold $p=0.5$.
}
\label{fig:results}
\end{figure*}

With the optimized 16-channel LU-invariant representation, EntNet reaches essentially perfect performance on the $N=16$ XY benchmark, reaching 100\% AUROC (Fig.\,\ref{fig:results}d). Moreover, EntNet achieves a true-negative rate (TNR) of 0.997 on an independent benchmark of 2,000 fully separable states with nontrivial classical correlations (Fig.\,\ref{fig:results}g), demonstrating its ability to distinguish entanglement from classical correlations.
A more stringent test is whether the same model generalizes beyond the conditions seen during training. We therefore test extrapolation to different Hamiltonian families, higher-order certificates, and larger system sizes. Unless otherwise stated, all results below directly use the identical EntNet model trained only on $N=16$ separable states and \EcertTwo{} states.

We first examine transfer across different Hamiltonian families. On the XXZ, frustrated $J_1$--$J_2$, and random-local-Hamiltonian benchmarks, EntNet achieves near-perfect AUROC and accurately identifies higher-order certified \EcertThree{} and \EcertFour{} states (Fig.\,\ref{fig:results}c).
The main performance degradation occurs in the $J_1$--$J_2$ \EcertTwo{} sector, where recall decreases to 0.86 while AUROC remains 1.0000.
This indicates a limitation associated with the default decision threshold of $p=0.5$, rather than a failure to correctly rank entangled states relative to separable controls. Further analysis indicates that this deficit arises from
insufficient coverage of relevant spatial correlation patterns in the training distribution, rather than missing information in the selected features. To verify this, we augment the training set with independently generated states exhibiting strong entanglement between sites separated by two lattice spacings ($r=2$). Without changing the selected 16 channels or including any $J_1$--$J_2$ states in training, the \EcertTwo{} recall on the $J_1$--$J_2$ benchmark is restored to 1.0000 (Appendix~D).

We further evaluate EntNet on two independent $N=32$ benchmarks, comprising finite-temperature XY-chain states and synthetically constructed states with diverse entanglement structures. Both benchmarks include states whose entanglement can only be certified from three- or four-site reduced density matrices. 
Without retraining or feature reselection, EntNet maintains near-perfect ranking and high recall across these different entanglement classes [Fig.~\ref{fig:results}(e,f)]. These results indicate that the learned representation remains effective for larger systems and extends beyond the specific quantum states and entanglement certificates used in training. 
This size transfer is consistent with the use of RoPE, which encodes relative spatial positions without restricting the model to the training sequence length. 
Self-attention allows the model to combine correlation information from sites at different separations, while RoPE makes their relative displacement explicit in the attention weights. The same positional relationships can therefore be transfered to longer chains without introducing new position-specific parameters.

Together, these results demonstrate that EntNet learns transferable correlation signatures associated with many-body entanglement, rather than merely fitting the training distribution, with the learned representation generalizing across Hamiltonian families, system sizes, and entanglement certificate levels.

\subsection{Role of LU invariance in the input representation}
We next examine the role of enforcing LU symmetry directly in the input representation of EntNet. We compare three treatments of the same physical observables: raw Pauli correlators with no LU constraint, raw correlators trained with a soft consistency loss under random local basis rotations, and the exact LU-invariant representation used by the 16-channel EntNet. 
The training data and Transformer architecture are otherwise kept as similar as possible, with implementation details provided in Appendix~C.

The difference becomes pronounced once we deploy the trained EntNet variants and apply them under transferability tests in Fig.\,\ref{fig:results}d-g. 
The model based on raw correlators generalizes poorly and can exhibit substantial false-positive rates, while the soft LU constraint improves performance but remains less stable across benchmarks. By contrast, the exact LU-invariant representation used in this work preserves both strong recall of certified entangled states and high true-negative rates for constructively separable states across system size and certificate order. 
These results highlight the effectiveness of incorporating known physical symmetries into the representation, providing a robust foundation for entanglement detection.

\subsection{Finite-temperature entanglement and learned physical boundaries}
We finally test whether the learned classifier behaves consistently along continuous finite-temperature trajectories, where the underlying physics can be understood analytically. 
For a fixed $N=16$ XY Hamiltonian with the thermal state $\rho(T)=e^{-H/T}/Z$, writing $\beta=1/T$ and $d=2^N$, at the high-temperature limit the thermal state can be expanded as
\begin{equation}
\rho(T)=\frac{I}{d}-\frac{1}{dT}\left(H-\frac{\mathrm{Tr}H}{d}I\right)+\mathcal{O}(T^{-2}).
\label{eq:highT-expansion}
\end{equation}
Therefore,
\begin{equation}
\lim_{T\to \infty}\rho(T)=\frac{I}{2^N}=\bigotimes_{i=1}^{N}\frac{I_i}{2},
\end{equation}
which is an exactly separable state. Thus, regardless of the low-temperature behavior, every thermal trajectory must eventually approach the same separable infinite-temperature limit.

\begin{figure*}[t]
\centering
\includegraphics[width=0.95\textwidth]{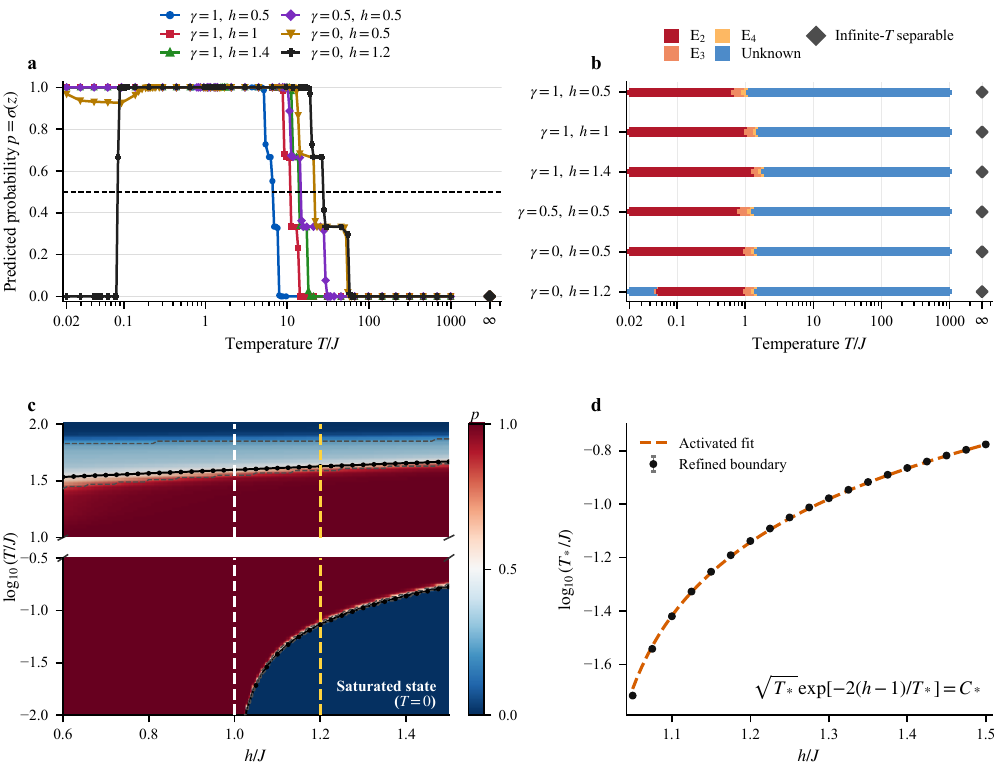}
\caption{\textbf{Thermal response and activated reentry in EntNet predictions.}
\textbf{a.} EntNet predictions $p=\sigma(z)$ along six finite-temperature trajectories of the $N=16$ XY chain, using the original trained ensemble. The dashed line marks the classification threshold $p=0.5$, and diamonds indicate the common infinite-temperature limit.
\textbf{b.} Entanglement certificates along the same trajectories, independently determined from PPT and CCNR tests on two- to four-site reduced density matrices. ``Unknown'' indicates the absence of an entanglement certificate, rather than separability, while the infinite-temperature state is exactly separable.
\textbf{c.} EntNet predictions in the $(h,T)$ plane of the XX chain ($\gamma=0$), using an ensemble trained with additional states exhibiting strong distance-two entanglement. For $h/J>1$, thermal excitations drive the initially separable ground state into a region classified as entangled, before the predicted probability decreases again at higher temperatures. Black contours mark $p=0.5$, while gray bands indicate variations across independently trained models. The temperature axis contains a break separating the low- and high-temperature regions.
\textbf{d.} Refined lower boundary $T^*(h)$ of the model-positive region and its fit to the thermally activated form $\sqrt{T^*}\exp[-2(h-1)/T^*]=C^*$, with energies expressed in units of $J$. The fit uses 19 boundary points over $1.05\leq h/J\leq1.50$, yielding $C^*=1.041\times10^{-3}$ and $R^2_{\log_{10}T^*}=0.9993$.
This boundary characterizes the learned classifier and should not be interpreted as a rigorous
entanglement threshold or a thermodynamic phase transition.
}
\label{fig:xy-thermal}
\end{figure*}

This trend is reproduced directly by the EntNet predictions on the XY Hamiltonian in Eq.\eqref{eq:hamxy}. Along six representative thermal trajectories, the classifier output $p=\sigma(z)$ approaches zero at high temperature and converges toward the common infinite-temperature limit (Fig.\,\ref{fig:xy-thermal}a). 
In particular, for $\gamma=0$ and $h=1.2$, EntNet exhibits a reentrant classification pattern with increasing temperature, transitioning from separable to entangled and back to separable predictions.
To resolve the underlying physical evolution of these same trajectories, we independently evaluate their PPT/CCNR entanglement certificates (Fig.\,\ref{fig:xy-thermal}b). The states evolve through \EcertTwo{}, \EcertThree{}/\EcertFour{}, and eventually uncertified regimes as thermal mixing increases. This certificate-resolved analysis provides a direct physical interpretation of the continuous changes seen in the EntNet score, showing that its predictions track the evolving entanglement structure of the thermal states.

Moreover, Fig.\,\ref{fig:xy-thermal}c shows the classifier output as a heatmap over the full $(h,T)$ plane in the XX limit, $\gamma=0$, where the temperature-dependence of entanglement is non-monotonic. 
For $h>1$, the zero-temperature ground state is fully polarized and therefore separable, and EntNet correctly assigns this region $p<0.5$. Upon heating, however, thermally populated excited states introduce nontrivial correlations into the Gibbs state, causing the classifier to enter a model-positive region with $p>0.5$. 
At even higher temperature, thermal mixing suppresses these correlations and drives the state back toward the maximally mixed separable limit. The resulting evolution represents a thermally induced reentrant classifier response  and is consistent with known finite-temperature entanglement behavior in cyclic XX chains \cite{CanosaRossignoli2007}.

The lower decision boundary of EntNet admits a simple physical interpretation in terms of thermally activated Jordan--Wigner fermions. In the Jordan--Wigner representation \cite{LiebSchultzMattis1961,BarouchMcCoyDresden1970}, the $\gamma=0$ quasiparticle dispersion is (setting $J=1$)
\begin{equation}
\epsilon_k=2(h-\cos k)\simeq2(h-1)+k^2,\qquad h>1,\quad k\simeq0.
\label{eq:xx-low-energy}
\end{equation}
The excitation gap is therefore $\Delta=2(h-1)$. At low temperature, the density of thermally excited Jordan-Wigner fermions scales as
\begin{equation}
n_{\rm th}\propto\int dk\,e^{-\epsilon_k/T}\propto\sqrt{T}\,e^{-2(h-1)/T}.
\label{eq:xx-activated-density}
\end{equation}
In the dilute-excitation regime, local spin correlators acquire corrections controlled by this thermally activated quasiparticle density. It is therefore natural to expect the lower classifier boundary to occur when these excitation-induced correlations reach an approximately fixed sensitivity scale $C_*$. This gives
\begin{equation}
\sqrt{T_*}\,\exp\!\left[-\frac{2(h-1)}{T_*}\right]=C_*,
\label{eq:activated-boundary}
\end{equation}
where $T_*(h)$ denotes the lower $p=0.5$ decision boundary.
As shown in Fig.\,\ref{fig:xy-thermal}d, this activated form closely follows the refined classifier boundary over $1.05\leq h\leq1.50$. An unweighted fit in $\log T_*$ gives $C_*=1.041\times10^{-3}$, $R^2_{\log_{10}T_*}=0.9993$, and $\mathrm{RMSE}(\log_{10}T_*)=0.0068$. The close agreement is consistent with the sensitivity of EntNet's classifier to the excitation-gap scale that controls thermally populated Jordan--Wigner fermions. 
EntNet therefore learns a decision boundary across the $(h,T)$ plane that quantitatively follows the quasiparticle activation law. 
Importantly, this field-dependent decision boundary of EntNet should not be interpreted as a rigorous separability boundary. 
In the XX limit ($\gamma=0$), for example, the exchange Hamiltonian commutes with the total $Z$ magnetization, $[H_{\rm XX}(0),\sum_i Z_i]=0$. At any fixed $T>0$, Gibbs states at different finite fields are related by invertible local filtering, $\rho_h(T)\propto F_h\rho_0(T)F_h$, where $F_h=\bigotimes_i\exp[hZ_i/(2T)]$. 
Since invertible local transformation preserves full separability, the exact separability property at fixed positive temperature is independent of $h$ \cite{RossignoliCanosa2005,CanosaRossignoli2006}. 
The field-dependent lower contour therefore characterizes EntNet's sensitivity to thermally activated correlation signatures rather than an exact onset of entanglement.
The resulting prediction map connects three distinct regimes: an initially separable polarized ground state, an intermediate-temperature region where thermally excited correlations produce a positive classifier response, and the high-temperature separable limit.
The reentrant structure and thermally activated scaling above show that EntNet captures finite-temperature signatures of many-body entanglement directly from low-order observables. 
These results suggest that the continuous prediction score of EntNet carries qualitative information about the evolving entanglement structure, beyond its role as a binary classification indicator.

\section{Discussion}
We have developed EntNet to identify signatures of many-body entanglement from the collective spatial organization of low-order spin correlations. 
Rather than relying on a single or a few predefined observables and entanglement witnesses, as done in conventional scattering probes, EntNet combines information from diverse spin correlations across different orders and spatial separations, uncovering complementary signatures of many-body entanglement.
With only 16 LU-invariant feature channels, EntNet detects states certified by two- to four-site entanglement criteria and generalizes across system sizes and Hamiltonian families with strong performance.
At finite temperature, EntNet further exhibits a reentrant classification pattern, with its lower decision boundary quantitatively following the quasiparticle activation scale set by the excitation gap of Jordan-Wigner fermions. 
These results establish that a compact set of low-order observables can encode transferable, physically-informed signatures of many-body entanglement across correlation order, spatial scale, system size, and Hamiltonian family.

Several limitations define the next steps for EntNet. First, the present framework is developed for one-dimensional spin chains, and extending the representation to two-dimensional and three-dimensional lattices and more complex geometries will be important for broader many-body applications. 
Second, EntNet currently focuses on spin degrees of freedom; extending the same symmetry-aware strategy to fermionic systems would test whether a similar strategy can remain valid beyond spin Hamiltonians. 
Third, although the final representation is compact, each LU-invariant channel is constructed from multiple underlying Pauli correlators, making measurement efficiency an important consideration for actual experimental implementation. 
Programmable quantum simulators, such as superconducting-qubit arrays and trapped-ion systems, offer a promising route, where two-, three-, and four-site Pauli correlators can be estimated through repeated state preparation, local basis rotations, and projective measurements \cite{Karamlou2024,Brydges2019}. 
For quantum materials, scattering techniques offer complementary possibilities: neutron scattering probes two-spin correlations \cite{scheie2022quantum}, polarized neutron scattering can access three-spin chiral correlations under suitable conditions \cite{grigoriev2005critical}, and resonant inelastic X-ray scattering (RIXS) is sensitive to selected higher-order correlations \cite{van2007theory,braicovich2009dispersion}. However, connecting these scattering observables to multiple local equal-time correlators required by EntNet will require further development.
Reducing the required measurement set while preserving classification accuracy will be essential for translating EntNet to these experimental settings.

Nevertheless, the EntNet framework opens a route toward scalable entanglement detection in many-body systems where full state tomography is impractical and only restricted sets of correlation measurements are experimentally accessible \cite{GuhneToth2009,friis2019entanglement,elben2023randomized}. 
Conventional entanglement witnesses provide rigorous certification when an appropriate inequality is available, whereas EntNet offers a complementary way to screen and characterize many-body states from a compact set of low-order correlation functions \cite{friis2019entanglement}. 
This is especially useful for scalable characterization of engineered entanglement. Potential applications include quantum simulators \cite{islam2015measuring}, superconducting-qubit platforms \cite{gold2021entanglement}, and strongly correlated quantum materials \cite{scheie2024proximate,hales2023witnessing}, where many-body entanglement is routinely generated, manipulated, or inferred from restricted measurements. 
Such models could help map regions of detectable entanglement in phase diagrams, prioritize experimentally prepared states for stronger certification, and guide the selection of informative correlation measurements.
More broadly, this approach suggests a paradigm shift from applying predefined entanglement witnesses toward discovering informative combinations of experimentally accessible correlations, potentially guiding the development of more efficient and transferable entanglement detection protocols.


\bibliography{references}

\makeatletter
\close@column@grid
\makeatother
\input{supplement}

\end{document}

%% file: supplement.tex
\appendix
\small

\section*{Appendix A: Label construction and benchmark scope}
\label{app:labels}

\subsection{Certificate hierarchy}

Every negative example is constructive. A separable density operator is generated as
\begin{equation}
  \rho_{\Sep}=\sum_{\mu=1}^{M}w_\mu
  \bigotimes_{i=1}^{N}\rho_i^{(\mu)},
  \qquad w_\mu\geq0,\quad \sum_\mu w_\mu=1 .
  \label{eq:app-sep}
\end{equation}
Consequently, a negative label does not mean merely ``not detected by a witness.'' For a bipartition $A|\bar A$, a state is separable across that cut when it is a convex mixture of products $\rho_A\otimes\rho_{\bar A}$. In a product basis $\{|i\rangle_A|\alpha\rangle_{\bar A}\}$, partial transposition on $A$ is defined by
\begin{equation}
 \langle i\alpha|\rho^{T_A}|j\beta\rangle
 =\langle j\alpha|\rho|i\beta\rangle .
 \label{eq:partial-transpose}
\end{equation}
Separable states across the cut have $\rho^{T_A}\succeq0$ \cite{Peres1996,Horodecki1996}; a negative eigenvalue therefore certifies entanglement. The converse need not hold in generic multipartite dimensions. For the complementary computable cross-norm or realignment (CCNR) test, write $\rho_{i\alpha,j\beta}$ for matrix elements of a bipartite state and rearrange them as
\begin{equation}
 [\mathcal R(\rho)]_{ij,\alpha\beta}=\rho_{i\alpha,j\beta}.
 \label{eq:realignment}
\end{equation}
Separable states obey $\|\mathcal R(\rho)\|_1\leq1$; a larger trace norm is a sufficient entanglement certificate \cite{Rudolph2005,ChenWu2003}. We test the inequivalent bipartitions of each reduced subsystem using fixed numerical tolerances. Positive labels are assigned by scanning these reduced states through a fixed hierarchy. In the datasets used here, \EcertTwo{} denotes a state first certified by a negative partial transpose of a two-site marginal. \EcertThree{} and \EcertFour{} denote states for which no lower-order certificate is found but a three- or four-site marginal is certified by the partial-transpose or CCNR criterion. A state that is not certified by this hierarchy is marked ``unknown'' and excluded from all binary metrics. Thus the classifier is trained against certified positives and constructive negatives; it is never rewarded for treating an uncertified state as separable.

Finite-temperature states are mixed because of the Gibbs distribution, and the synthetic training families also contain purification- and mixture-induced mixedness. This physical mixedness is part of the task. By contrast, the input features reported here are exact expectation values. Finite sampling, readout errors, and imperfect state preparation have not yet been included, so the present results should not be described as experimentally noise robust.

\subsection{Datasets and split roles}

\begin{table}[h]
\centering
\caption{Principal datasets used for training, feature selection, and evaluation. Unknown labels are excluded from binary metrics and never treated as separable.}
\label{tab:data}
\small
\begin{tabular}{llll}
\toprule
Dataset & Size & Composition & Role\\
\midrule
Primary training & 35,000 & 17,500 Sep + 17,500 \EcertTwo{} & classifier training\\
Feature-selection data & 2,812 & 1,406 Sep/Ent pairs & channel ranking\\
Distance-two support & 1,400 & 700 Sep/\EcertTwo{} pairs & training control\\
$N=16$ XY & 912 & 432 Sep, 432 Ent, 48 unknown & evaluation\\
$N=32$ XY & 500 & 20 Sep, 432 Ent, 48 unknown & size transfer\\
$N=32$ synthetic states & 2,000 & 400 Sep, 1,600 Ent & transfer\\
Independent separable states & 2,000 & constructive mixtures & false-positive control\\
\bottomrule
\end{tabular}
\end{table}

\begin{table}[h]
\centering
\caption{Composition of the independent $N=16$ Gibbs-state benchmarks from different Hamiltonian families. Adaptive temperature refinement targets regions near the disappearance of the two-site entanglement certificate. No artificial noise is added; unknown states are excluded from performance metrics.}
\label{tab:crossham-data}
\small
\begin{tabular}{lrrrrrr}
\toprule
Family & Total & Sep & \EcertTwo{} & \EcertThree{} & \EcertFour{} & Unknown\\
\midrule
XXZ & 343 & 79 & 103 & 52 & 30 & 79\\
$J_1$--$J_2$ & 463 & 30 & 155 & 80 & 30 & 168\\
Random local $H$ & 626 & 39 & 238 & 87 & 40 & 222\\
\bottomrule
\end{tabular}
\end{table}

The datasets serve distinct purposes. The channel ranking is obtained from the primary training distribution together with two auxiliary collections of matched separable and entangled states. After this ranking is frozen, the final 16-channel EntNet classifiers and the other top-$K$ classifiers are initialized and trained independently using only the 35,000 primary training states. A separate control supplements this training distribution with states exhibiting strong distance-two correlations, without changing the selected channels. Parent states, matched pairs, and their local-unitary copies remain within a single data split. No evaluation benchmark is used for feature selection or classifier training.

The positive-label provenance is important for interpreting generalization across certificate levels. Both the primary training distribution and the additional distance-two training distribution contain only \EcertTwo{} positive labels. Neither training route contains \EcertThree{} or \EcertFour{} labels, and no higher-order-certified state from any evaluation benchmark is used for training or model selection.

\subsection{Training-set generation}

\textit{Primary training distribution.---}
The full parent dataset contains 50,000 $N=16$ states, split 70:15:15 within each label. Only its 35,000 training states enter the primary classifier-training route. Its 25,000 constructive negative examples comprise 5,000 product states, 17,500 convex mixtures of 2--32 product states, and 2,500 classical GHZ mixtures. More explicitly, every negative has a known decomposition
\begin{equation}
 \rho_{\Sep}=\sum_{\mu=1}^{K}w_\mu
 \bigotimes_{i=1}^{N}\frac{I+\bm r_{\mu i}\!\cdot\!\bm\sigma}{2}.
 \label{eq:dv1-separable-construction}
\end{equation}
Local Bloch vectors have isotropic directions and a mixture of pure and Bloch-ball radii. The product family has $K=1$, whereas the general family mixes 2--32 product components with sampled weights. The classical GHZ mixture is not a coherent GHZ state; it is the explicitly separable state
\begin{equation}
 \rho_{\mathrm{cGHZ}}=\frac12\bigotimes_i\rho(+\bm n_i)
 +\frac12\bigotimes_i\rho(-\bm n_i),
 \qquad \rho(\bm n)=\frac{I+\bm n\!\cdot\!\bm\sigma}{2},
 \label{eq:classical-ghz-mixture}
\end{equation}
with independently sampled local axes. Finally, independent Haar-distributed single-qubit rotations are applied at all sites. These operations preserve the displayed convex product decomposition, so the negative label is known constructively rather than inferred from failure of an entanglement criterion.

The 25,000 positive parent states are random matrix product states (MPSs) with requested bond dimensions between 2 and 16, followed by local dephasing, depolarizing, or amplitude-damping noise. Noise strengths range from easily certified states to states near certificate boundaries, which are located separately for each parent and noise channel. Every accepted positive retains a two-site negative-partial-transpose (NPT) marginal at separation at most two. Thus all positive examples used for the primary classifier training are \EcertTwo{} states.

\textit{Auxiliary feature-selection data.---}
Two independent collections of matched separable and entangled states are used only to rank candidate features. The first contains pairs constructed from product states and finite-depth entangling circuits, or from classical versus coherent GHZ and cluster blocks. The second pairs exactly separable ground states with entangled ground states generated from random-local Hamiltonians of the form in Eq.~\eqref{eq:random-local-hamiltonian}. Each collection contains 1,000 parent pairs, split by pair identity before feature selection. Their training partitions contain 706 and 700 pairs, respectively, giving the 2,812 states listed in Table~\ref{tab:data}. Neither auxiliary collection enters the final compact-classifier training. The random-local thermal benchmark uses independent Hamiltonian realizations, although its Hamiltonian family is also represented in the auxiliary feature-selection data. Its transfer result is therefore zero-shot for the final classifier, but the entire feature-selection pipeline is not completely unfamiliar with that Hamiltonian family.

\textit{Distance-two training support.---}
To test the effect of missing spatial patterns in the training distribution, we construct an independent collection of matched states from the $N=16$ open-boundary Interleaved-XXZ Hamiltonian in Eq.~\eqref{eq:interleaved-xxz}. This model is designed to exhibit strong correlations between sites separated by two lattice spacings. For each sampled Hamiltonian, two-site density-matrix renormalization group (DMRG) produces an \EcertTwo{} ground state whose most negative two-site partial-transpose eigenvalue occurs at separation two. The matched negative is an exactly polarized product ground state, established using the bondwise saturation bound
\begin{equation}
 h_{\rm sat}=J_2(\Delta_2+1)+\epsilon J_1(\Delta_1+1),
 \label{eq:r2support-hsat}
\end{equation}
with $|h|=h_{\rm sat}+\delta h$ and $\delta h>0$. The full collection contains 1,000 \EcertTwo{} states and 1,000 constructively separable states. Parent-wise 70:15:15 splitting yields 1,400 training, 300 validation, and 300 test states, without sharing parents across splits. The DMRG states satisfy an energy-variance-per-site check below $10^{-8}$. To characterize their spatial structure, we define
\begin{equation}
 R_{21}=\frac{\langle s_2[T^{(2)}(i,i+2)]\rangle_i}
 {\langle s_2[T^{(2)}(i,i+1)]\rangle_i+10^{-9}},
 \label{eq:r21}
\end{equation}
where the averages run over valid lattice anchors. Among the entangled states, 94.7\% satisfy $R_{21}>3$, confirming that their distance-two correlations dominate the corresponding nearest-neighbor correlations. The $J_1$--$J_2$ evaluation states and classifier predictions are not used to construct or filter this additional dataset. The separable partners supplement, rather than replace, the more diverse separable families in the primary training distribution.

\subsection{Hamiltonians and XY parameter-space structure}

The XY Hamiltonian is given in Eq.~\eqref{eq:hamxy}. The open-chain transfer benchmarks use $S_i^\alpha=\sigma_i^\alpha/2$ and the Hamiltonians
\begin{align}
 H_{\rm XXZ}&=J\sum_{i=1}^{N-1}
 (S_i^xS_{i+1}^x+S_i^yS_{i+1}^y+\Delta S_i^zS_{i+1}^z)
 -h\sum_{i=1}^{N}S_i^z,
 \label{eq:xxz-hamiltonian}\\
 H_{J_1J_2}&=J_1\sum_{i=1}^{N-1}\bm S_i\!\cdot\!\bm S_{i+1}
 +J_2\sum_{i=1}^{N-2}\bm S_i\!\cdot\!\bm S_{i+2}
 -h\sum_{i=1}^{N}S_i^z.
 \label{eq:j1j2-hamiltonian}
\end{align}
The additional distance-two training states are generated from a distinct Interleaved-XXZ family,
\begin{align}
 H_{\rm IXXZ}={}&J_2\sum_{i=1}^{N-2}
 \bigl(S_i^xS_{i+2}^x+S_i^yS_{i+2}^y
 +\Delta_2S_i^zS_{i+2}^z\bigr)\notag\\
 &+\epsilon J_1\sum_{i=1}^{N-1}
 \bigl(S_i^xS_{i+1}^x+S_i^yS_{i+1}^y
 +\Delta_1S_i^zS_{i+1}^z\bigr)
 -h\sum_{i=1}^{N}S_i^z .
 \label{eq:interleaved-xxz}
\end{align}
At small $\epsilon$, its distance-two bonds form two weakly coupled even and odd XXZ subchains. These equations specify the signs and operator normalizations; the use of Pauli operators in the XY model versus spin operators in the open-chain models matters when comparing field scales.

The XY benchmarks use periodic boundary conditions and the parameter region $0\leq\gamma\leq1$, $0\leq h\leq1.5$. The single-particle dispersion is
\begin{equation}
 \varepsilon_k=2\sqrt{(h-\cos k)^2+\gamma^2\sin^2 k},
 \label{eq:xy-dispersion}
\end{equation}
which reveals the gap-closing lines and the following zero-temperature structure \cite{LiebSchultzMattis1961,BarouchMcCoyDresden1970,BarouchMcCoy1971}.

For $0<\gamma\leq1$ and $0\leq h<1$, the thermodynamic ground state belongs to the $\mathbb Z_2$-ordered ferromagnetic phase. On a finite periodic chain, Hamiltonian eigenstates may be chosen with definite parity; away from exact degeneracy, such a state does not itself select either broken-symmetry orientation. The thermodynamic Ising transition lies at $h=1$ for $\gamma>0$. The line $\gamma=0$ gives the number-conserving XX model, which is gapless for $0\leq h<1$ and fully polarized for $h>1$.

The circle
\begin{equation}
 \gamma^2+h^2=1
 \label{eq:factorization-circle}
\end{equation}
is not a phase boundary. Instead, it is the factorization locus within the ordered region \cite{Kurmann1982}, where the ground space admits two symmetry-breaking product states
\begin{equation}
 |\Phi_{\pm}\rangle=\bigotimes_{i=1}^{N}
 \left(\cos\phi\,|0\rangle_i\pm\sin\phi\,|1\rangle_i\right),\qquad
 \phi=\frac12\arccos\!\left(\frac{h}{1+\gamma}\right).
 \label{eq:factorized-state}
\end{equation}
Their local Bloch vectors are $(\pm\sin2\phi,0,\cos2\phi)$. Definite-parity ground states can instead be formed from even and odd superpositions of these product branches. This finite-size symmetry issue does not change the existence of the exact separable branches.

For $h>1$, the anisotropic XY chain is in the field-polarized paramagnetic phase. However, this region should not be identified entirely with an exactly saturated product eigenstate. At finite field and $\gamma>0$, the anisotropic exchange contains pair-flip terms, so the ground state remains weakly dressed and approaches full polarization only as $h\to\infty$. The exact saturated state $|0\rangle^{\otimes N}$ used as a benchmark anchor occurs on the XX line $\gamma=0$ with $h>1$, where the magnetization-conserving exchange annihilates the all-up state. For $T>0$, the evaluated object is the Gibbs mixed state $\rho_T=e^{-H_{\rm XY}/T}/\operatorname{Tr}e^{-H_{\rm XY}/T}$ rather than an individual eigenstate. Phase labels therefore describe the zero-temperature spectrum underlying the sampled thermal crossover.

The open XXZ benchmark uses Eq.~\eqref{eq:xxz-hamiltonian} with Latin-hypercube coverage $J\in\{-1,+1\}$, $-2.5\leq\Delta\leq3.0$, and $-3.5\leq h\leq3.5$, supplemented by structured anchors including $J=0$ product controls. The open $J_1$--$J_2$ benchmark uses Eq.~\eqref{eq:j1j2-hamiltonian} with $J_1=1$, $-0.5\leq J_2\leq1.5$, and $-4\leq h\leq4$. Positive $J_2$ frustrates the antiferromagnetic nearest-neighbor exchange. The empirical condition $J_2/J_1\geq0.9$ isolates a difficult strongly frustrated subset for analysis; it is not presented as a phase boundary. In both families, large-field fully polarized states are included as constructively separable controls whenever a bondwise positive-semidefinite bound establishes that they are ground states.

The random-local Hamiltonian family uses an open chain of Pauli spins,
\begin{equation}
 H_{\rm RL}=-\sum_{i=1}^{N} f_i\,\bm n_i\!\cdot\!\bm\sigma_i
 +\sum_{i=1}^{N-1}\sum_{a,b=x,y,z}
 \bigl[-c_i n_i^a n_{i+1}^b+\lambda A_i^{ab}\bigr]
 \sigma_i^a\sigma_{i+1}^b .
 \label{eq:random-local-hamiltonian}
\end{equation}
Here each $\bm n_i$ is a unit vector, while $A_i$ is a real bond matrix orthogonal to $\bm n_i\bm n_{i+1}^{\mathsf T}$ and normalized to unit Frobenius norm. The sampled fields are sufficiently strong that the commuting $\lambda=0$ parent has a proved product ground state. Independent, weakly homogeneous, and smoothly varying site profiles broaden the local axes and couplings, while nonzero $\lambda$ introduces interactions that do not commute with the local fields. The exact profile distributions and sampling seeds are recorded in the generation code. The evaluation states are independent of the random-local ground-state pairs used in feature selection and do not enter final classifier training.

For the distance-two support family, we fix $J_1=J_2=1$ and vary the anisotropies, field, and weak nearest-neighbor coupling. Its dominant distance-two bonds connect the even and odd sublattices into two XXZ backbones, whereas the $\epsilon$-scaled nearest-neighbor term weakly couples them. Entangled fields are sampled symmetrically from fixed fractions up to 0.78 of the saturation bound in Eq.~\eqref{eq:r2support-hsat}; separable partners are placed beyond that bound. This family is deliberately distinct from the isotropic frustrated $J_1$--$J_2$ evaluation model and provides additional distance-two correlation patterns without reproducing the evaluated Hamiltonians.

\subsection{Evaluation-set generation}

\textit{XY benchmarks.---}
The $N=16$ XY benchmark consists of 480 finite-temperature Gibbs states sampled across the parameter region, including neighborhoods of the critical line, factorization circle, and low-temperature limit, together with 432 analytically constructed separable states. Of these separable states, 20 are product-state anchors covering the factorization circle and saturated XX line, and the remaining 412 are independently constructed factorized states equally divided between the two branches in Eq.~\eqref{eq:factorized-state}. Exact free-fermion methods provide the correlators. All energies and temperatures are expressed in units of the XY exchange coefficient. Two-site marginals and contiguous three- and four-site marginals are screened in hierarchical order. The final benchmark contains 432 separable states, 313 \EcertTwo{} states, 119 higher-order-certified states, and 48 unknown states. The unknown states are retained for diagnostic purposes but excluded from binary evaluation metrics.

The $N=32$ XY benchmark uses the same 480 Hamiltonian parameter points and 20 analytic product anchors as the corresponding $N=16$ sample, but recomputes all observables and entanglement certificates at the larger system size. It contains 20 separable states, 312 \EcertTwo{} states, 80 \EcertThree{} states, 40 \EcertFour{} states, and 48 unknown states. It therefore tests application of the frozen network to longer chains rather than reusing the $N=16$ feature vectors.

\textit{Synthetic $N=32$ benchmark and separable controls.---}
The synthetic benchmark contains 2,000 independently constructed states, including 400 constructively separable states and 1,600 entangled states distributed across the three certificate levels. No dense $2^{32}$ density matrix is formed. Each embedded-block entangled state takes the form
\begin{equation}
 \rho=U_{B}\rho_{B}U_{B}^{\dagger}\otimes
 \bigotimes_{i\notin B}\rho_i,
 \qquad |B|\in\{2,3,4\},
 \label{eq:dpi-embedded-block}
\end{equation}
where $B$ is a randomly placed contiguous block, $U_B$ is a tensor product of independently sampled single-qubit Haar rotations, and the remaining sites form a random product background. Strict negatives are generated as convex mixtures of 32-site product states. Thus the required moments and reduced density matrices can be evaluated from the small embedded blocks and product backgrounds without constructing a full-chain density matrix.

Higher-order entangled blocks are generated from GHZ, graph, and shallow-circuit states, with noise applied before certificate screening. Three elementary noise maps are used:
\begin{align}
 \mathcal W_p(\rho)&=(1-p)\rho+pI/2^{|B|},\notag\\
 \mathcal D_p^{\otimes |B|}(\rho)&=
 \sum_{\bm a}K_{\bm a}\rho K_{\bm a}^{\dagger},\quad
 \{K_a\}=\{\sqrt{1-p}I,\sqrt p|0\rangle\langle0|,
                 \sqrt p|1\rangle\langle1|\},\notag\\
 \mathcal A_p^{\otimes |B|}(\rho)&=
 \sum_{\bm a}A_{\bm a}\rho A_{\bm a}^{\dagger},\quad
 A_0=\begin{pmatrix}1&0\\0&\sqrt{1-p}\end{pmatrix},\quad
 A_1=\begin{pmatrix}0&\sqrt p\\0&0\end{pmatrix}.
 \label{eq:dpi-noise-channels}
\end{align}
Here the same $p$ is applied independently to every qubit in the embedded block. This denotes a tensor product of local quantum channels acting on a possibly entangled density operator, not a factorization of the density operator itself. The \EcertThree{} sources include a white-noised three-qubit GHZ state, a weakly white-noised GHZ state mixed toward a product endpoint, and GHZ-derived states subjected to white noise, dephasing, or amplitude damping. The \EcertFour{} sources include white-noised four-qubit star-graph states, dephased and weakly white-noised four-qubit GHZ states, and shallow random circuits. The \EcertTwo{} sources include embedded Bell pairs, depth-1--3 random two-qubit circuits, and entangled two-site Heisenberg Gibbs states. The resulting benchmark contains 400 states of each of the separable and \EcertTwo{} classes, together with 600 \EcertThree{} and 600 \EcertFour{} states.

Every candidate is screened in the fixed order \EcertTwo{} $\rightarrow$ \EcertThree{} $\rightarrow$ \EcertFour{}. A two-site NPT certificate takes priority. An \EcertThree{} candidate must have no certified pair but must contain a certified contiguous triple. An \EcertFour{} candidate must have no lower-order certificate but must contain a certified contiguous quadruple. Fixed positive safety margins exclude numerical boundary cases. Candidates that fail the certificate or margin checks are not admitted to the benchmark. Separately, 2,000 additional constructively separable 32-site states are generated independently to assess false positives.

\textit{Cross-Hamiltonian benchmarks.---}
For the XXZ, $J_1$--$J_2$, and random-local Hamiltonian families, Gibbs states are sampled over temperature and the sampling is refined near the disappearance of two-site entanglement certificates. This procedure captures states whose entanglement is detectable only in higher-order marginals. Their class compositions are reported in Table~\ref{tab:crossham-data}. Each family contains at least ten independent parent Hamiltonians within each higher-order certificate class. No artificial noise channel is applied to these Gibbs states, and uncertified states remain unknown rather than becoming negative examples. The random-local thermal trajectories are independent of the random-local ground-state pairs used for feature selection. Hamiltonian parameters are sampled independently of classifier predictions; certificate transitions guide only temperature refinement. Evaluation states do not enter training or checkpoint selection. Classifier weights, normalization, and the decision threshold remain frozen throughout evaluation, with temperature grids and random seeds retained in the generation code.

\subsection{Finite-temperature correlators and convergence checks}

The XXZ Gibbs states are computed through matrix-product-state purification \cite{Verstraete2004,FeiguinWhite2005}. At $\beta=0$, each physical spin is maximally entangled with an ancilla,
\begin{equation}
 |I\rangle=\bigotimes_{i=1}^{N}
 \frac{|0\rangle_{p_i}|0\rangle_{a_i}+|1\rangle_{p_i}|1\rangle_{a_i}}
 {\sqrt2}.
 \label{eq:infinite-temperature-purification}
\end{equation}
Imaginary-time evolution acts only on the physical degrees of freedom,
\begin{equation}
 |\Psi_\beta\rangle=e^{-\beta H/2}|I\rangle,
 \qquad
 \rho_\beta=\frac{\operatorname{Tr}_{a}|\Psi_\beta\rangle
 \langle\Psi_\beta|}{\langle\Psi_\beta|\Psi_\beta\rangle}.
 \label{eq:thermal-purification}
\end{equation}
Temperatures are traversed from high to low using two-site imaginary-time time-dependent variational principle (TDVP) evolution with open boundaries \cite{Haegeman2016}. The required reduced density matrices are contracted directly from the purified MPS. Independent bond-dimension and time-step resolutions must agree on the certificate class, and the relevant reduced states and certificate margins must satisfy the convergence tolerances recorded in the implementation. Disagreement leaves a state uncertified.

The frustrated $J_1$--$J_2$ chain is evaluated using canonical thermal typicality \cite{SugiuraShimizu2013}, because purification becomes more expensive in the parameter regime considered. For a random phase vector $|r\rangle$ and $\beta=1/T$, Lanczos--Krylov propagation \cite{JaklicPrelovsek1994} approximates
\begin{equation}
 |\phi_r(\beta)\rangle=e^{-\beta H/2}|r\rangle .
 \label{eq:thermal-typical-vector}
\end{equation}
For each required support $S$, the reduced density matrix is estimated directly as
\begin{equation}
 \widehat\rho_S(\beta)=
 \frac{\sum_{r=1}^{R}\operatorname{Tr}_{\bar S}
 |\phi_r(\beta)\rangle\langle\phi_r(\beta)|}
 {\sum_{r=1}^{R}\langle\phi_r(\beta)|\phi_r(\beta)\rangle}.
 \label{eq:thermal-typical-rdm}
\end{equation}
Independent random-vector estimates and their pooled estimate must assign the same certificate order. We additionally require the certificate margin to exceed five times its block-estimated sampling uncertainty; otherwise the state is marked unknown.

For the random-local family, zero-temperature states are obtained by two-site DMRG and checked using energy variance. The initial finite-temperature scan uses the reduced-density-matrix-first thermal-typicality procedure above, while additional thermal trajectories use two independently resolved MPS purifications. Certification requires the corresponding stochastic or two-resolution agreement checks. Failed candidates remain unknown, not negative. The open boundaries and Pauli-operator normalization follow Eq.~\eqref{eq:random-local-hamiltonian}; no Hamiltonian coefficient or temperature is supplied to the classifier.

Finally, the Pauli tensors supplied to the invariant pipeline are evaluated from the corresponding reduced states,
\begin{equation}
 T^{(q)}_{a_1\cdots a_q}(i_1,\ldots,i_q)
 =\operatorname{Tr}\!\left[\rho_{i_1\cdots i_q}
 \sigma_{a_1}\otimes\cdots\otimes\sigma_{a_q}\right].
 \label{eq:thermal-pauli-from-rdm}
\end{equation}
Thus EntNet does not receive MPS tensors, Lanczos vectors, energies, or Hamiltonian parameters. It receives the same low-order local-unitary-invariant correlator features used for all other datasets.

\section*{Appendix B: Invariant feature construction}
\label{app:features}

\subsection{Why unfolding spectra are LU invariant}

For a $q$-site Pauli correlation tensor, independent on-site unitary transformations induce orthogonal rotations $R_i\in\mathrm{SO}(3)$. After grouping tensor indices into a bipartition $A|B$, the unfolded matrix transforms as
\begin{equation}
  T_{A|B}'=R_A T_{A|B}R_B^{\mathsf T},
  \qquad
  R_A=\bigotimes_{i\in A}R_i,
  \quad R_B=\bigotimes_{j\in B}R_j .
  \label{eq:app-unfold}
\end{equation}
Both $R_A$ and $R_B$ are orthogonal. Consequently, $(T_{A|B}')^{\mathsf T}T_{A|B}'$ is orthogonally similar to $T_{A|B}^{\mathsf T}T_{A|B}$, and its singular values are unchanged. Pair features use both the raw two-body tensor $T^{(2)}$ and the connected covariance
\begin{equation}
 C^{(2)}_{ab}(i,j)=T^{(2)}_{ab}(i,j)-m_i^a m_j^b,
 \qquad m_i^a=\mathrm{Tr}(\rho\sigma_i^a).
 \label{eq:connected}
\end{equation}
The subtraction removes the disconnected product of one-site Bloch vectors. Both terms transform covariantly, so $C^{(2)}(i,j)\mapsto R_iC^{(2)}(i,j)R_j^{\mathsf T}$, and its singular values are also exact local-unitary (LU) invariants rather than approximate invariants learned through data augmentation.

The measurement support includes pairs at separations $r=1,\ldots,4$, but only contiguous three- and four-site windows that fit inside the chain. It does not wrap at the feature boundary, even when the XY Hamiltonian itself has periodic boundary conditions. The final 16-channel representation uses pair separations $r=1$ and $r=4$, together with contiguous higher-order windows. Its 16 scalar invariants are computed from Pauli correlators; they are not 16 independently measured observables. The actual experimental measurement cost depends on how the underlying correlators are grouped and how measurement shots are allocated.

\subsection{Candidate dictionary and selected channels}

Let $s_k[X]$ denote the $k$th singular value of a matrix $X$, ordered nonincreasingly. The complete translation-tied candidate dictionary is defined as follows. The one-body block contains the Bloch-vector norm $\|\bm m_i\|_2$. For each pair distance $r=1,\ldots,4$ and each $X\in\{T^{(2)},C^{(2)}\}$, the two-body block contains
\begin{equation}
 \left\{s_1[X],s_2[X],s_3[X],\det X,
 \bm m_i^{\mathsf T}X\bm m_j,
 \|X\bm m_j\|_2,\|X^{\mathsf T}\bm m_i\|_2\right\}.
 \label{eq:app-pair-dictionary}
\end{equation}
This yields 14 pair invariants per distance and 56 in total. For a contiguous triple, we use both the raw tensor $T^{(3)}$ and the connected third cumulant
\begin{align}
 K^{(3)}_{abc}(i,j,k)={}&T^{(3)}_{abc}(i,j,k)
 -m_i^aT^{(2)}_{bc}(j,k)-m_j^bT^{(2)}_{ac}(i,k)\notag\\
 &-m_k^cT^{(2)}_{ab}(i,j)+2m_i^am_j^bm_k^c .
 \label{eq:app-connected-three}
\end{align}
For each tensor, the candidate dictionary contains its Frobenius norm and the three singular values of each of its three $1|2$ unfoldings, yielding 20 three-body invariants. Finally, the raw contiguous four-body tensor contributes its Frobenius norm, the three singular values of each of four $1|3$ unfoldings, and the nine singular values of each of the $12|34$, $13|24$, and $14|23$ unfoldings, giving 40 four-body invariants. No connected four-body tensor is included in this candidate pool. Together with the one-body norm, these quantities define 117 candidate physical channels. Geometry masks are deterministic inputs rather than selectable observables. Table~\ref{tab:channels} lists the selected 16 channels in consensus order.

\begin{table}[!htbp]
\centering
\caption{The 16 selected LU-invariant channels, ordered by consensus rank. For $T^{(3)}$, mode $a$ denotes the $a|(bc)$ unfolding. Four-body partitions refer to the site indices within a contiguous block.}
\label{tab:channels}
\small
\setlength{\tabcolsep}{5pt}
\begin{tabular}{rllll}
\toprule
Rank & Order & Support & Tensor / unfolding & Invariant\\
\midrule
1  & $T^{(3)}$ & contiguous & raw, mode $0$ & $s_3$\\
2  & $T^{(2)}$ & $r=1$ & raw & $s_2$\\
3  & $T^{(4)}$ & contiguous & raw, $14|23$ & $s_3$\\
4  & $T^{(3)}$ & contiguous & raw, mode $1$ & $s_3$\\
5  & $T^{(4)}$ & contiguous & raw, $13|24$ & $s_2$\\
6  & $T^{(4)}$ & contiguous & raw, $13|24$ & $s_3$\\
7  & $C^{(2)}$ & $r=1$ & connected &
$\|(C^{(2)})^{\mathsf T}\bm m_i\|_2$\\
8  & $T^{(4)}$ & contiguous & raw, $2|134$ & $s_2$\\
9  & $T^{(3)}$ & contiguous & raw, mode $1$ & $s_2$\\
10 & $T^{(4)}$ & contiguous & raw, $14|23$ & $s_2$\\
11 & $T^{(2)}$ & $r=4$ & raw & $s_2$\\
12 & $T^{(4)}$ & contiguous & raw, $2|134$ & $s_3$\\
13 & $T^{(4)}$ & contiguous & raw, $12|34$ & $s_2$\\
14 & $C^{(2)}$ & $r=4$ & connected & $s_2$\\
15 & $T^{(2)}$ & $r=1$ & raw & $s_3$\\
16 & $T^{(3)}$ & contiguous & raw, mode $0$ & $s_2$\\
\bottomrule
\end{tabular}
\end{table}

The selected representation contains five pair, four three-body, and seven four-body channels. Its pair channels occur only at separations $r=1$ and $r=4$, with no explicit pair channel at $r=2$ or $r=3$. Five deterministic geometry masks are supplied separately from the physical channels.

\section*{Appendix C: Selection, training, and evaluation protocol}
\label{app:protocol}

\subsection{Sparse channel selection}

Feature selection is label supervised but translation tied: each of the 117 physical channel types has one scalar gate shared across every lattice anchor. Geometry masks are never gated. For channel $c$, the hard-concrete relaxation \cite{Louizos2018} is
\begin{align}
 s_c&=\operatorname{sigmoid}\!\left[
 \frac{\log u-\log(1-u)+\log\alpha_c}{\beta_{\rm hc}}\right],\qquad
 u\sim U(0,1),\notag\\
 g_c&=\operatorname{clip}_{[0,1]}
 \left[s_c(\zeta-\gamma)+\gamma\right],
 \label{eq:hard-concrete-gate}
\end{align}
with $\beta_{\rm hc}=2/3$, $\gamma=-0.1$, and $\zeta=1.1$. Its analytic nonzero probability is
\begin{equation}
 \pi_c=\operatorname{sigmoid}\!\left[
 \log\alpha_c-\beta_{\rm hc}\log(-\gamma/\zeta)\right],
 \qquad \widehat K=\sum_{c=1}^{N_{\rm total}}\pi_c .
 \label{eq:expected-channel-count}
\end{equation}
During selector training, the normalized channel values are multiplied by stochastic gates $g_c$. The objective is
\begin{equation}
 \mathcal L_{\rm sel}=\mathcal L_{\rm BCE}
 +\eta_e\left[
 0.01\frac{\widehat K}{N_{\rm total}}
 +20\left(
 \frac{\widehat K-N_{\rm target}}{N_{\rm total}}
 \right)^2\right],
 \qquad \eta_e=\min(1,e/25),
 \label{eq:app-selector-loss}
\end{equation}
where the 25-epoch ramp prevents the sparsity penalty from suppressing channels before the classifier has learned a useful signal. Here $N_{\rm total}=117$ and $N_{\rm target}=32$ for the reported selector. The soft target $N_{\rm target}$ stabilizes a complete ordering; it is not the final input width. After warm-up, checkpoints are selected by minimizing validation binary cross entropy (BCE) plus $0.02|\widehat K-N_{\rm target}|/N_{\rm target}$.

Five independent selectors are trained on the feature-selection data described in Appendix~A. Thus the frozen channel ordering is not obtained solely from the primary classifier-training distribution. Within each run, channels are ranked according to their learned $\log\alpha_c$. The consensus order is determined first by mean rank across the five selectors, then by mean $\log\alpha_c$, and finally by the fixed channel index. Truncating this ranking gives nested channel budgets. After the ranking is frozen, all selector gates, selector parameters, and classifier weights used during feature selection are discarded. Each compact classifier is initialized and trained independently on the selected physical channels. EntNet uses the first 16 channels in the consensus order, together with all geometry masks. No $N=32$ evaluation example enters gate learning, ranking, normalization, checkpoint selection, or early stopping.

As a provenance diagnostic, repeating the five-selector procedure using only the primary training and validation states produces a top-16 list containing nine of the original 16 channels. This alternative ordering is not substituted into any reported classifier. All results use the same frozen channel identities and ordering.

\subsection{EntNet Transformer architecture}

At anchor $i$, the 16 selected invariant values are concatenated with five deterministic availability masks: four for pair separations $r=1,\ldots,4$ and one for contiguous three-site support. With open feature boundaries, a contiguous four-site support is valid at exactly the same anchors as an $r=3$ pair, so the corresponding pair mask also specifies four-site validity. Invalid physical entries are zero padded before token assembly. A separate binary sequence mask $V_{bj}\in\{0,1\}$ distinguishes real tokens from batch padding when chains of different lengths are processed together; the classification token is always valid. These masks contain only geometry information, not entanglement labels.

The resulting tensor $X^{(0)}\in\mathbb R^{B\times N\times21}$ is standardized channelwise using training-split statistics and passed through a shared $21\rightarrow128$ affine projection. A learned classification (CLS) token is prepended, giving a Transformer input of shape $[B,N+1,128]$. Each of the four Transformer layers contains four attention heads, each with width $d_h=32$. Rotary positional encoding (RoPE) \cite{Su2024} acts on paired coordinates of queries and keys,
\begin{equation}
 \widetilde{\bm q}_{p,\ell}=R(p\omega_\ell)\bm q_{p,\ell},\qquad
 \widetilde{\bm k}_{p,\ell}=R(p\omega_\ell)\bm k_{p,\ell},\qquad
 \omega_\ell=10{,}000^{-2\ell/d_h},
 \label{eq:rope-rotation}
\end{equation}
where
\begin{equation}
 R(\theta)=
 \begin{pmatrix}
 \cos\theta&-\sin\theta\\
 \sin\theta&\cos\theta
 \end{pmatrix}.
 \label{eq:rope-matrix}
\end{equation}
Values are not rotated. Orthogonality gives
\begin{equation}
 \widetilde{\bm q}_{i,\ell}^{\mathsf T}
 \widetilde{\bm k}_{j,\ell}
 =\bm q_{i,\ell}^{\mathsf T}R[(j-i)\omega_\ell]\bm k_{j,\ell},
 \label{eq:rope-relative}
\end{equation}
so the query--key phase depends on relative displacement without introducing a learned absolute-position vector into the physical features.

For one attention head, the masked attention operation is
\begin{equation}
 \operatorname{Attn}(Q,K,V)=
 \operatorname{softmax}\!\left(
 \frac{\widetilde Q\widetilde K^{\mathsf T}}{\sqrt{32}}
 +M_{\rm attn}\right)V,
 \label{eq:masked-rope-attention}
\end{equation}
where $(M_{\rm attn})_{bqk}=0$ for valid keys and $-\infty$ for padded keys in the mathematical idealization. The implementation uses the smallest finite attention-logit value in place of $-\infty$ before softmax; the valid CLS key ensures that every softmax remains defined. Attention weights and block outputs at padded query positions are explicitly reset to zero. Each pre-normalized residual block is therefore
\begin{align}
 Y^{(\ell)}&=X^{(\ell)}+
 \operatorname{MHA}[\operatorname{LN}(X^{(\ell)})],\notag\\
 X^{(\ell+1)}&=Y^{(\ell)}+
 \operatorname{FFN}[\operatorname{LN}(Y^{(\ell)})],
 \label{eq:transformer-block}
\end{align}
with a $128\rightarrow192\rightarrow128$ GELU feed-forward network and dropout 0.1. After four blocks, only the CLS representation is read. A final LayerNorm and $128\rightarrow128\rightarrow1$ GELU classification head produce the logit $z$ and probability $p=\operatorname{sigmoid}(z)$. Each compact model contains 483,969 trainable parameters. The architecture allows each anchor to compare its local invariant features with patterns elsewhere in the chain while keeping the physical input exactly LU invariant.

Each compact checkpoint minimizes unweighted BCE-with-logits using AdamW \cite{LoshchilovHutter2019}. Three independently initialized classifiers use the same optimizer settings, validation-BCE checkpoint rule, and early-stopping procedure. Selector runs use separate seeds and a sparsity-penalty warm-up; their checkpoints are discarded after the consensus ranking is frozen. Learning rates, scheduler settings, and random seeds are recorded in the training configuration.

\subsection{Raw-correlator LU controls}

We compare EntNet with two Transformers trained on exactly the same 35,000 primary training states and the same parent-safe training, validation, and test partitions. Their tokens contain 147 Cartesian Pauli-correlation components: three one-body components, four sets of nine two-body components at separations $r=1,\ldots,4$, 27 contiguous three-body components, and 81 contiguous four-body components. Together with the same five deterministic geometry masks, this gives an input width of 152. All physical components required to construct the selected EntNet invariants are present in the raw representation. Unavailable boundary components are zero padded, and their masks are retained. Channel means and standard deviations are computed only from valid anchors in the primary training split, with masks left unstandardized.

Both controls use the same four-layer, four-head, pre-normalized RoPE Transformer architecture, hidden width 128, feed-forward width 192, CLS pooling, optimizer, early-stopping criterion, and three random seeds as EntNet. They also use the same two-layer $128\rightarrow128\rightarrow1$ classification head. Only the input projection width differs; the raw-input model has more trainable parameters because it receives a larger set of physical channels.

The first control is trained using canonical raw correlators and unweighted binary cross entropy. The second imposes a soft LU-consistency constraint: the same network processes a canonical input and a second input obtained by independently rotating the local basis at every site using Haar-random single-qubit unitaries. These rotations act on every index of the one- through four-body Pauli tensors, while geometry masks remain unchanged. Its loss is
\begin{equation}
 \mathcal L_{\rm Siamese}=\tfrac12\bigl[\mathcal L_{\rm BCE}(z,y)
 +\mathcal L_{\rm BCE}(z',y)\bigr]
 +\lambda_{\rm LU}\bigl|\sigma(z)-\sigma(z')\bigr|,
 \qquad \lambda_{\rm LU}=0.5,
 \label{eq:siamese-control-loss}
\end{equation}
averaged over each minibatch. This encourages basis consistency but does not make the raw representation exactly invariant. Checkpoints for both controls are selected solely by validation BCE on canonical inputs. Evaluation benchmarks are used only after training, with the same three-seed probability averaging as EntNet. No test-dependent checkpoint selection is performed.

The raw-versus-soft comparison holds the input information and architecture fixed, testing the combined effect of rotation-augmented supervision and the soft consistency penalty; it does not separately isolate these two contributions. The comparison with EntNet instead tests an analytically invariant and compressed representation against the full raw tensor representation. Although the latter contains the information required to construct the selected invariants, the representation dimensionality and inductive bias also differ. Therefore, any observed generalization gap cannot be attributed exclusively to LU invariance. Per-seed metrics and individual predictions are retained in the experiment records; the soft-control recall on the synthetic $N=32$ benchmark varies substantially across its three seeds.

\begin{table}[h]
\centering
\caption{Comparison of input representations using correlations through fourth order. All models are trained on the same primary $N=16$ training distribution. Predictions are averaged as probabilities across three independently initialized classifiers. Recall and false-positive rate (FPR) use the fixed threshold $p=0.5$. For the independent separable benchmark, only FPR is defined.}
\label{tab:lu-input-control}
\small
\begin{tabular}{llcccc}
\toprule
Model & Benchmark & AUROC & \EcertTwo{} recall & \Egt{} recall & FPR\\
\midrule
Raw correlators, $T_{1-4}$ & $N=16$ XY & 0.6944 & 0.8914 & 1.0000 & 1.0000\\
 & $N=32$ XY & 0.3931 & 1.0000 & 1.0000 & 1.0000\\
 & $N=32$ synthetic & 0.1212 & 0 & 0 & 0.0200\\
 & Independent Sep & --- & --- & --- & 0.0600\\
\addlinespace
Soft LU, $T_{1-4}$ & $N=16$ XY & 0.5975 & 0.6326 & 0.8739 & 1.0000\\
 & $N=32$ XY & 0.6072 & 0.7468 & 0.9500 & 1.0000\\
 & $N=32$ synthetic & 0.6856 & 0.4075 & 0.5033 & 0.3325\\
 & Independent Sep & --- & --- & --- & 0.2885\\
\addlinespace
EntNet & $N=16$ XY & 1.0000 & 1.0000 & 1.0000 & 0\\
 & $N=32$ XY & 1.0000 & 1.0000 & 1.0000 & 0\\
 & $N=32$ synthetic & 0.9998 & 0.9875 & 0.9833 & 0\\
 & Independent Sep & --- & --- & --- & 0.0030\\
\bottomrule
\end{tabular}
\end{table}

On the synthetic $N=32$ benchmark, the soft constraint increases AUROC from 0.1212 to 0.6856 relative to the matched raw-correlator model, but also increases the false-positive rate from 0.0200 to 0.3325. Its false-positive rate on the independent separable benchmark rises from 0.0600 to 0.2885. The three-seed \EcertTwo{} recall for the soft-control model has a mean of 0.4133 and a sample standard deviation of 0.3400, indicating substantial variation across initializations. EntNet achieves the strongest combined discrimination and separable-state rejection in this comparison. This advantage is consistent with the benefits of exact LU invariance, but cannot isolate them from the effects of feature compression and representation design.

\subsection{Ensemble evaluation and benchmark metrics}

For retained checkpoint $s$, let $z_s(\bm x)$ denote the output logit and $p_s(\bm x)=\sigma[z_s(\bm x)]$ its predicted probability. The three-checkpoint ensemble averages probabilities and converts the result back to a logit for visualization:
\begin{equation}
 \bar p(\bm x)=\frac13\sum_{s=1}^{3}p_s(\bm x),\qquad
 z(\bm x)=\log\frac{\bar p(\bm x)}{1-\bar p(\bm x)}.
 \label{eq:ensemble-score}
\end{equation}
Let $\mathcal E=\EcertTwo{}\cup \EcertThree{}\cup \EcertFour{}$ denote the certified entangled set and $\mathcal S$ the constructively separable set. At the fixed threshold $\bar p=0.5$, class-specific recall and false-positive rate are defined by
\begin{align}
 \operatorname{Recall}_{E_q}
 &=\frac{1}{|E_q|}\sum_{\bm x\in E_q}\mathbf 1[z(\bm x)\geq0],\notag\\
 \operatorname{FPR}
 &=\frac{1}{|\mathcal S|}\sum_{\bm x\in\mathcal S}
 \mathbf 1[z(\bm x)\geq0].
 \label{eq:threshold-metrics}
\end{align}
The area under the receiver operating characteristic curve (AUROC) is the threshold-independent probability that a randomly selected certified entangled state scores above a randomly selected separable state, assigning half weight to ties:
\begin{align}
 \operatorname{AUROC}=\frac{1}{|\mathcal E||\mathcal S|}
 \sum_{\bm x\in\mathcal E}\sum_{\bm y\in\mathcal S}
 \bigl[&\mathbf1\{s(\bm x)>s(\bm y)\}\notag\\
 &+\tfrac12\mathbf1\{s(\bm x)=s(\bm y)\}\bigr].
 \label{eq:auroc-definition}
\end{align}
For the ensemble results reported here, $s=\bar p$ is the arithmetic mean of three predicted probabilities, not an average of logits. Unknown states are excluded from these metrics but retained in diagnostic score plots. The decision threshold is fixed before evaluation.

To evaluate different feature budgets, we freeze the 117-channel consensus ranking and train a new classifier independently for each nested top-$K$ prefix, with $K=4,8,\ldots,32$. Every budget uses the same primary training and validation partitions, three seeds, architecture, optimizer, and validation-BCE checkpoint rule; only the number of selected physical channels changes. All five geometry masks remain. Evaluation uses the $N=16$ XY benchmark containing 432 constructively separable states, 313 \EcertTwo{} states, 119 higher-order-certified states, and 48 unknown states. Unknown states are excluded from all performance metrics. Table~\ref{tab:k-budget} reports the three-seed probability-averaged ensemble results at the fixed threshold $\bar p=0.5$.

The reference 16-channel model preceded completion of the full budget scan, but the $N=16$ XY results were considered when adopting $K=16$. Therefore, this scan is a retrospective comparison rather than an independent prospective test of an entirely prespecified channel count. It was not used to rerank candidate channels or select individual checkpoints, and the cross-Hamiltonian and $N=32$ benchmarks remained outside the budget comparison. The input width $K+5$ counts numerical model coordinates, not independent Pauli measurement settings, because multiple spectral invariants can be computed from the same correlation tensor.

\begin{table}[h]
\caption{Channel-budget comparison on the $N=16$ XY benchmark. Classifiers use nested prefixes of the same frozen channel ranking and are independently trained on the primary training distribution. Values are computed from three-seed probability-averaged predictions. FPR refers to 432 constructively separable controls.}
\label{tab:k-budget}
\small
\setlength{\tabcolsep}{2pt}
\begin{ruledtabular}
\begin{tabular}{cccccc}
$K$ & Width & AUROC & \EcertTwo{} recall & \Egt{} recall & FPR\\
\hline
4  & 9  & 0.6667 & 0.6709 & 0.1513 & 0\\
8  & 13 & 0.6968 & 0.7764 & 0.1597 & 0\\
12 & 17 & 0.8981 & 0.7316 & 0.7731 & 0\\
16 & 21 & 1.0000 & 1.0000 & 1.0000 & 0\\
20 & 25 & 0.9606 & 0.9201 & 0.9496 & 0\\
24 & 29 & 0.9954 & 0.9521 & 0.9328 & 0\\
28 & 33 & 0.7905 & 0.9042 & 0.8655 & 1.0000\\
32 & 37 & 0.7245 & 0.9105 & 0.8992 & 1.0000\\
\end{tabular}
\end{ruledtabular}
\end{table}

For this benchmark, the independently trained $K=16$ ensemble reproduces the four aggregate performance metrics of the established EntNet ensemble, although the checkpoints and individual predicted probabilities differ. Performance does not improve monotonically with $K$: the $K=28$ and $K=32$ models classify all 432 constructively separable controls as entangled at the fixed threshold, even though their AUROC remains above chance. Additional selected features therefore do not guarantee better generalization or decision-threshold behavior. These retrospective results support the present compact representation but do not establish a universally optimal channel count.

\begin{table}[h]
\caption{Performance of the frozen EntNet ensemble on the $N=16$ Gibbs-state benchmarks from different Hamiltonian families. Counts in parentheses denote Sep/\EcertTwo{}/\EcertThree{}/\EcertFour{}. Unknown states are excluded from all metrics. Recall and FPR are evaluated at the fixed threshold $p=0.5$.}
\label{tab:crossham}
\scriptsize
\setlength{\tabcolsep}{1.6pt}
\begin{ruledtabular}
\begin{tabular}{lccccc}
Family (Sep/\EcertTwo{}/\EcertThree{}/\EcertFour{}) & AUROC & FPR & \EcertTwo{} & \EcertThree{} & \EcertFour{}\\
\hline
XXZ (79/103/52/30) & 1.0000 & 0 & 0.9903 & 1.0000 & 1.0000\\
$J_1$--$J_2$ (30/155/80/30) & 1.0000 & 0 & 0.8645 & 1.0000 & 1.0000\\
Random local $H$ (39/238/87/40) & 1.0000 & 0 & 1.0000 & 1.0000 & 1.0000\\
\end{tabular}
\end{ruledtabular}
\end{table}

The system-size transfer tests apply the same $N=16$-trained network to $N=32$ token sequences without updating its parameters. The variable-length architecture accepts the longer chains directly. Table~\ref{tab:lu-input-control} reports the model's performance on both the $N=32$ XY states and the independently constructed synthetic states. A separate field-of-view intervention isolating the contribution of sites 17--32 has not been verified for this checkpoint; consequently, the present results establish successful evaluation at larger $N$ but do not quantify how strongly the network uses information from the additional sites.

\section*{Appendix D: Controlled interventions}
\label{app:ablations}

\subsection{Diagnosing missing spatial correlation patterns}

The lower \EcertTwo{} recall on the $J_1$--$J_2$ benchmark reflects a mismatch between the original training distribution and the spatial correlations encountered in this Hamiltonian family. In the regime $J_1\rightarrow0$, the even and odd sites form nearly independent chains, making correlations at separation $r=2$ particularly important. Although EntNet contains no explicit $r=2$ pair channel, its contiguous three- and four-body features probe correlations spanning this distance. The observed deficit could therefore arise either from information missing in the selected representation or from insufficient exposure to these patterns during training.

To distinguish these explanations, we augment the training distribution with independently constructed states exhibiting strong distance-two entanglement, while keeping the same 16 input channels. This restores positive predictions for all 155 \EcertTwo{} states in the $J_1$--$J_2$ benchmark, without introducing any $J_1$--$J_2$ evaluation state into training. The recovery demonstrates that the selected representation contains enough information to recognize these states and identifies insufficient training coverage, rather than the absence of a dedicated $r=2$ pair observable, as the main source of the original deficit.

\begin{figure*}[t]
\centering
\includegraphics[width=0.8\textwidth]{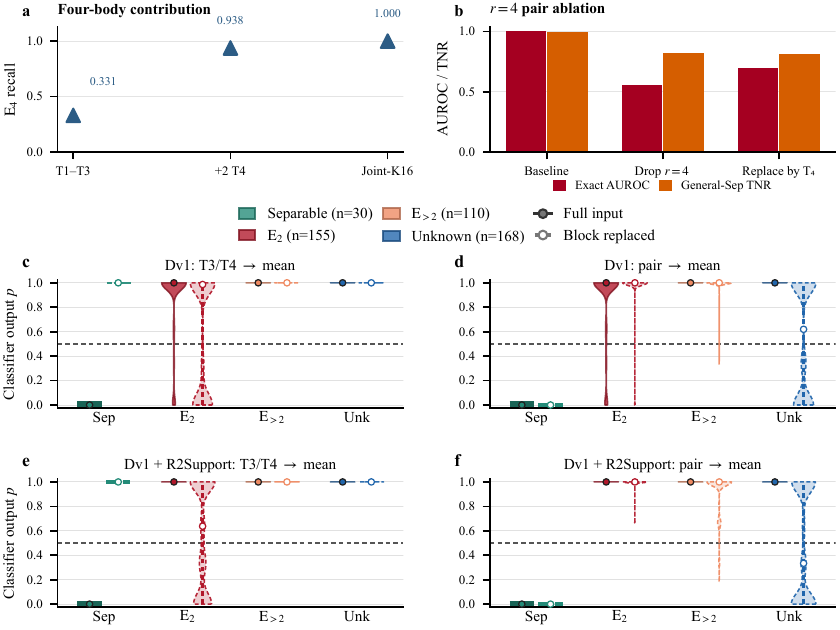}
\caption{\textbf{Complementary contributions of pair and higher-order correlations.}
\textbf{(a)} \EcertFour{} recall on the synthetic $N=32$ benchmark using features through third order, after adding two selected four-body channels, and with the complete 16-channel representation.
\textbf{(b)} Changes in exact-state AUROC and independent separable-state true-negative rate (TNR) after removing the selected $r=4$ pair channels or replacing them with additional four-body invariants.
\textbf{(c)--(f)} Classifier-score distributions for $N=16$ $J_1$--$J_2$ Gibbs states before and after replacing selected feature blocks with their training-set means. Three- and four-body channels are replaced in (c,e), while pair channels are replaced in (d,f). Panels (c,d) use the original training distribution; panels (e,f) use the same distribution supplemented with independently generated states exhibiting strong distance-two correlations. Within each state class, the original and modified inputs are shown on the left and right, respectively. Circles indicate medians, vertical segments show interquartile ranges, and the dashed line marks $p=0.5$. Because modified inputs need not correspond to physical density matrices, these interventions measure the model's dependence on feature blocks rather than causal contributions to entanglement.
}
\label{fig:block-intervention}
\end{figure*}

To examine how EntNet combines correlations of different orders, we replace either the three-/four-body feature block or the pair-feature block with its training-set mean, leaving the remaining inputs unchanged. As shown in Fig.~\ref{fig:block-intervention}(c--f), replacing higher-order features degrades predictions even for \EcertTwo{} states, although their labels are defined by two-site NPT certificates. Conversely, replacing pair features also changes predictions for states whose entanglement is certified only at higher order. Thus the classifier does not simply assign distinct input blocks to the \EcertTwo{}, \EcertThree{}, and \EcertFour{} certificate levels.

Instead, EntNet uses the combined spatial and tensor structure of pair, three-body, and four-body correlations. Higher-order features provide information about how a pair-correlation pattern should be interpreted, while pair features supply complementary spatial information about higher-order structures. This behavior is consistent with learning joint signatures of entanglement rather than separately approximating the two-, three-, and four-site tests used to generate labels. Because replacing a feature block by its mean may produce a feature vector that does not correspond to any physical density matrix, these interventions diagnose model dependence and should not be interpreted as physically realizable changes to a quantum state.

Two further controls test whether the selected channels merely provide interchangeable versions of the same signal. Removing four-body features substantially reduces the recognition of \EcertFour{} states. Removing the selected $r=4$ pair channels also degrades performance, and replacing them with additional four-body invariants does not recover the loss [Fig.~\ref{fig:block-intervention}(a,b)]. These observations indicate that the selected representation benefits from complementary correlation orders and spatial length scales.

\subsection{$J_1$--$J_2$ feature-block interventions}

\begin{table}[h]
\centering
\caption{Numbers of $J_1$--$J_2$ states classified as entangled ($p\geq0.5$) before and after replacing a feature block by the corresponding training-set means. The higher-order-certified class contains 80 \EcertThree{} and 30 \EcertFour{} states. ``Original'' denotes training on the primary distribution alone; ``Augmented'' adds 1,400 independent states with strong distance-two correlation patterns, without changing the selected 16 channels.}
\label{tab:train-mean}
\small
\begin{tabular}{llrrrr}
\toprule
Training & Class & $n$ & Full & No $T^{(3)}/T^{(4)}$ & No pair\\
\midrule
Original & Sep & 30 & 0 & 30 & 0\\
& \EcertTwo{} & 155 & 134 & 81 & 153\\
& \Egt{} & 110 & 110 & 110 & 109\\
& Unknown & 168 & 168 & 168 & 84\\
\midrule
Augmented & Sep & 30 & 0 & 30 & 0\\
& \EcertTwo{} & 155 & 155 & 83 & 155\\
& \Egt{} & 110 & 110 & 110 & 106\\
& Unknown & 168 & 168 & 168 & 79\\
\bottomrule
\end{tabular}
\end{table}

The aggregate higher-order-certified counts conceal some dependence on certificate order. Replacing pair features leaves all 80 \EcertThree{} states above threshold in both models. For \EcertFour{} states, the number remaining above threshold changes from 30 to 29 with the original training distribution and from 30 to 26 with augmented training. In the latter case, the median \EcertFour{} probability decreases from 0.999999 to 0.671. The augmented model is trained on the 35,000 primary training states and 1,400 additional distance-two support states, with no $J_1$--$J_2$ or $N=32$ evaluation states. The complete per-state predictions underlying Fig.~\ref{fig:block-intervention} are retained with the figure data.

\subsection{Distance-two training-support control}

Here $r_{\rm cert}=2$ denotes a state whose most negative two-site partial-transpose eigenvalue occurs at separation two. The ratio $R_{21}$, defined in Eq.~\eqref{eq:r21}, compares spatially averaged second singular values of pair-correlation matrices at separations two and one. With the original 16 selected channels frozen, augmenting the training distribution with the independently generated distance-two states restores positive predictions for all 155 \EcertTwo{} states in the expanded $J_1$--$J_2$ evaluation set. No state from that evaluation set enters the augmented training distribution. Since the selected pair channels occur only at $r=1$ and $r=4$, while contiguous $T^{(3)}$ and $T^{(4)}$ windows span sites separated by two lattice spacings, this recovery shows that the existing representation can recognize the relevant spatial pattern once the classifier encounters suitable examples during training. It does not require introducing an additional input channel.

For the exact-state ablation in Fig.~\ref{fig:block-intervention}(b), the AUROC is evaluated on a 500-state subset of the $N=16$ XY benchmark, comprising 20 constructively separable states, 313 \EcertTwo{} states, 119 higher-order-certified states, and 48 unknown states. The unknown states are excluded from AUROC. The independent 2,000-state separable benchmark is used separately to evaluate the true-negative rate. These controls isolate the effects of specific selected feature groups without altering the classifier-training distribution.